\documentclass[conference]{IEEEtran}
\IEEEoverridecommandlockouts
\usepackage{cite}
\usepackage{amsmath,amssymb,amsfonts}
\usepackage{graphicx}
\usepackage{textcomp}
\usepackage{xcolor}
\usepackage{CJKutf8}  
\usepackage{subfigure} 
\usepackage{algorithm}
\usepackage{algorithmic}
\usepackage{makecell}
\usepackage{hyperref}
\hypersetup{hypertex=true,
            colorlinks=true,
            linkcolor=blue,  
            anchorcolor=blue,
            citecolor=blue}
\usepackage[all]{hypcap}  
\usepackage{tikz}
\usetikzlibrary{positioning}
\usetikzlibrary{math}
\usetikzlibrary{fpu}
\usetikzlibrary{arrows.meta}

\def\BibTeX{{\rm B\kern-.05em{\sc i\kern-.025em b}\kern-.08em
    T\kern-.1667em\lower.7ex\hbox{E}\kern-.125emX}}

\begin{document}
\begin{CJK}{UTF8}{gbsn}  

\title{A Novel Dual Predictors Framework of PEE\\
}

\author{\IEEEauthorblockN{1\textsuperscript{st} Fangjian Shen}
\IEEEauthorblockA{\textit{dept. name of organization (of Aff.)} \\
\textit{name of organization (of Aff.)}\\
City, Country \\
email address or ORCID}
\and
\IEEEauthorblockN{1\textsuperscript{st} Yicheng Zheng}
\IEEEauthorblockA{\textit{dept. name of organization (of Aff.)} \\
\textit{name of organization (of Aff.)}\\
City, Country \\
email address or ORCID}
\and
\IEEEauthorblockN{1\textsuperscript{st} Songyou Li}
\IEEEauthorblockA{\textit{dept. name of organization (of Aff.)} \\
\textit{name of organization (of Aff.)}\\
City, Country \\
email address or ORCID}
}

\maketitle

\begin{abstract}
This document is a model and instructions for \LaTeX.
This and the IEEEtran.cls file define the components of your paper [title, text, heads, etc.]. *CRITICAL: Do Not Use Symbols, Special Characters, Footnotes, 
or Math in Paper Title or Abstract.
\end{abstract}


\section{Introduction}
With the rapid development of information technology and the widespread use of digital media, the security and confidentiality of digital information have become increasingly important. Data hiding, as a covert communication technology that embeds secret information in digital media, has been applied to more and more fields \cite{bib_RDH_app-1, bib_RDH_app-2, bib_RDH_app-3}. However, in the domains of digital copyright protection, telemedicine diagnosis, and military communication, it is essential that the original media can be fully recovered without any loss of information after extracting the secret information. Therefore, reversible data hiding (RDH) has garnered significant attention in the research community \cite{bib_RDH-1, bib_RDH-2} and has witnessed rapid development in recent years.

Currently, the two primary methods for reversible data hiding are \textbf{difference expansion (DE)} and \textbf{histogram shifting (HS)}. DE achieves it by modifying the difference between neighboring pixel values in the image \cite{bib_DE}, but the bit per pixel (bpp) of this method is low and the distortion is large. Whereas the HS algorithm \cite{bib_HS-1, bib_HS-2, bib_HS-3} shifts the zero point of the histogram to the peak point and embeds confidential information into the corresponding bins. Nevertheless, the quality of the embedded image is constrained by the positional relationship between the peak and zero point. To address these limitations, \textbf{prediction-error expansion (PEE)} \cite{bib_C-PEE-1, bib_C-PEE-2, bib_C-PEE-3, bib_C-PEE-4} was proposed in subsequent research.

PEE employs a specific predictor to obtain the prediction value based on the context of each pixel. It then subtracts the prediction value from the original pixel value to derive prediction-error. By counting the frequency of the prediction error, \textbf{prediction error histogram (PEH)} is generated. Compared to the pixel histogram, PEH takes advantage of the similarity between image pixels. So PEH has higher peaks, which results in larger embedding capacity (EC) and less distortion. Different from HS, which shifts between peaks and zeros, PEE selects bins -1 and 0 for expansion while shifting the other bins accordingly to ensure reversibility.

However, the choice of bins -1 and 0 for expansion in PEE does not always guarantee minimum distortion when capacity is low. Moreover, the prediction accuracy tends to decrease on images with drastic gray-scale changes, leading to dispersed PEH where the peak bin may not be located at 0 or -1. To overcome these challenges, later research introduced \textbf{PEE with optimal expansion bins selection (O-PEE)} \cite{bib_O-PEE-1, bib_O-PEE-2}, which adaptively determines the expansion bins that minimize distortion while satisfying embedding capacity through exhaustive search, making it more effective than PEE. Additionally, \textbf{PEE with adaptive embedding (A-PEE)} \cite{bib_A-PEE-1, bib_A-PEE-2, bib_A-PEE-3, bib_A-PEE-4} introduces complexity to measure drastic content of each pixel's context changes. Since high complexity often results in high prediction error, A-PEE reduces distortion by filtering out pixels with high complexity, which makes the distribution of PEH sharper for less histogram shifts. Combining the advantages of both A-PEE and O-PEE, \textbf{PEE which combines both adaptive embedding and optimal expansion bins selection (AO-PEE)} \cite{bib_AO-PEE-1, bib_AO-PEE-2}  achieves superior results. Furthermore, pairwise PEE extends PEH to two-dimensional (2D) PEH by counting prediction-errors of pixel pairs on the diagonal of mutually disjoint $2 \times 2$ pixel blocks. It adaptively discards mapping directions with large distortion, which not only obtains good embedding performance but also garners high capacity.

Unlike AO-PEE, which directly filters out pixels with high complexity, in \cite{bib_MHM}, Li et al. proposed \textbf{multiple histogram modification (MHM)}, where complexity of each pixel is determined by summing both horizontal and vertical absolute differences of every two consecutive pixels in the context of each pixel. Then pixels will be divided into several intervals, and we generate PEH for each interval. Expansion bins are adaptively selected for each PEH, enabling efficient utilization of pixels with varying complexity levels. MHM can be considered a generalization of AO-PEE: when the histograms corresponding to high complexity classes remain unchanged and the expansion bins for low complexity ones are the same, MHM is equivalent to AO-PEE. Compared with AO-PEE, MHM significantly improved peak signal-to-noise ratio (PSNR). 

Afterwards, many researchers have made further improvements to MHM. \textbf{Optimal MHM}, proposed by Qi et al. in \cite{bib_optimal_MHM}, addresses the optimization problem of expansion bins selection by continuously reducing the original problem to smaller subproblems. In \cite{bib_high_capacity_MHM}, Ou et al. introduced \textbf{high-capacity MHM}, which implements high capacity embedding by selecting multiple pairs of expansion bins for each PEH. Additionally, in \cite{bib_PRE} proposed by Xiao et al. , the prediction residual is calculated by removing the rounding operation to replace the prediction-error, and \textbf{prediction residual histogram(PRH)} is derived. Then they designed a new mapping mechanism to generate more expansion bins based on the PRH. To achieve higher image quality when embedding secret message, MHM is combined with PRE to get MHM-PRE, which surpasses previous MHM-like algorithms in terms of image quality.

The sharpness of PEH determines the final image quality, but the distribution of the PEH is closely related to the performance of predictor. Consequently, numerous research efforts have focused on enhancing the accuracy of predictors. The most widely used predictor is the rhombus predictor \cite{bib_rhombor_double}, which has proven effective in generating sharp PEH to achieve a better embedding result. However, adopting only one predictor is not always reliable. The prediction capability of one single predictor is inevitably confined, accurately predicting certain points while introducing biases in the prediction for other points.

\begin{figure*}
    \centering
    \subfigure[2D-PEH of image lena generated by our proposed method]{\includegraphics[width=0.45\linewidth]{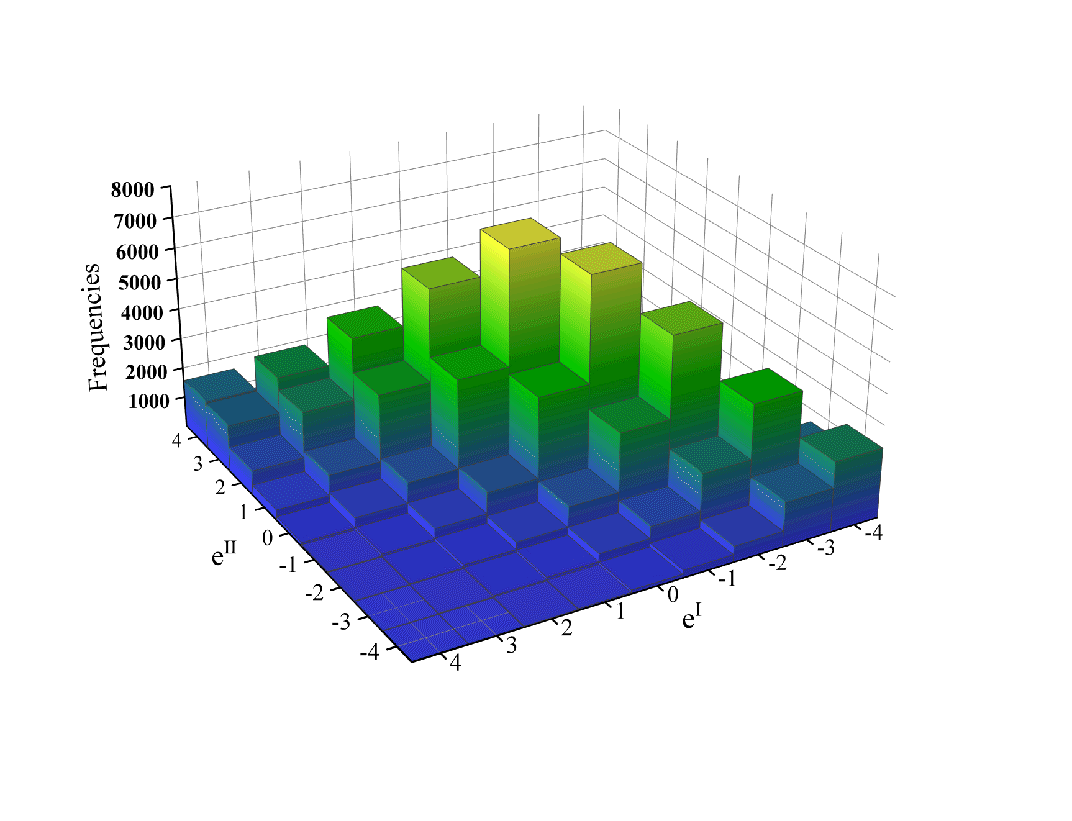}}
    \hspace{0.5cm}
    \subfigure[2D-PEH of image lena generated by pairwise-PEE]{\includegraphics[width=0.45\linewidth]{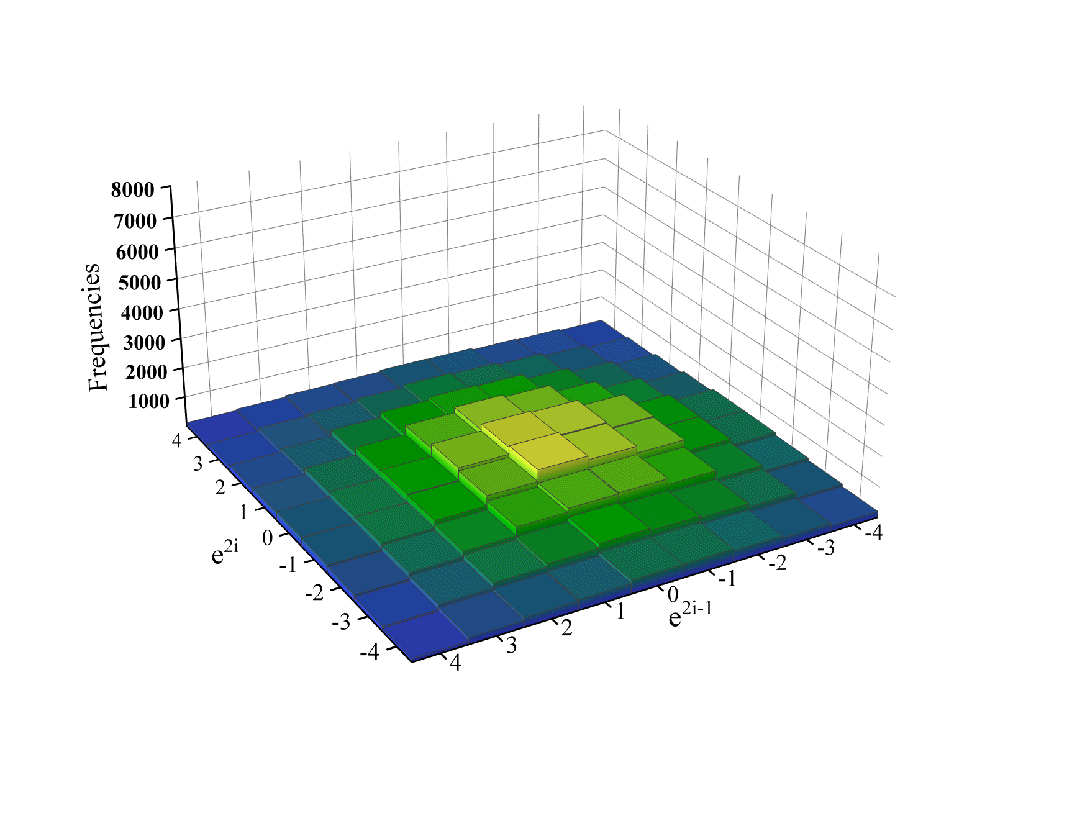}}
    \hspace{0.5cm}
    \subfigure[2D-PEH of image Baboon generated by our proposed method]{\includegraphics[width=0.45\linewidth]{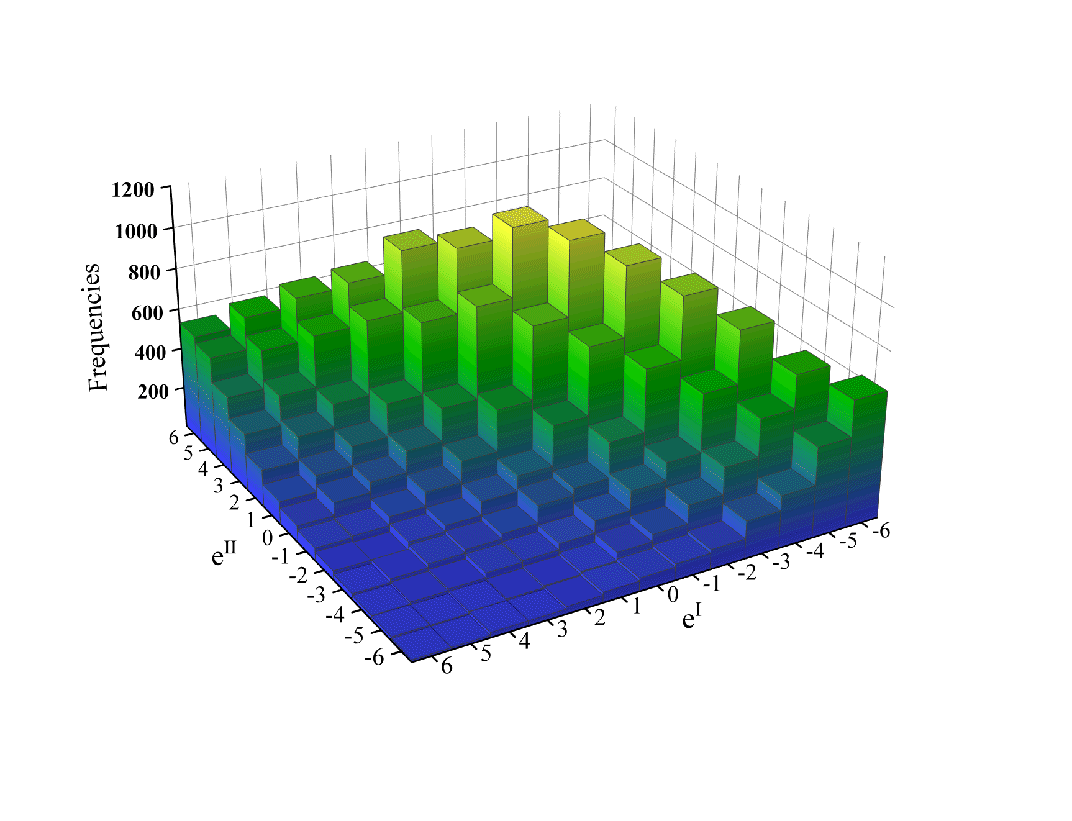}}
    \hspace{0.5cm}
    \subfigure[2D-PEH of image Baboon generated by pairwise-PEE]{\includegraphics[width=0.45\linewidth]{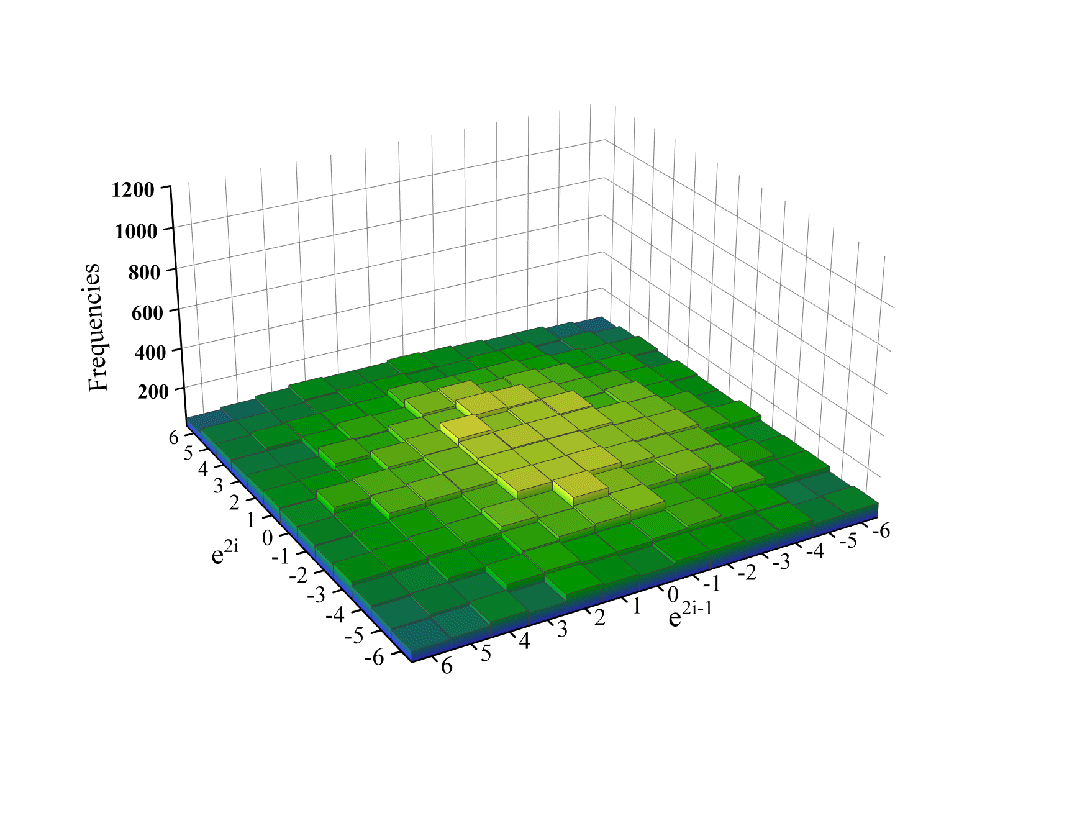}}
    
    \caption{2D-PEH comparison between the proposed method and C-PEE}
    \label{figure:hist-comparison}
\end{figure*}

In this paper, we propose an improved 2D-PEH by employing distinct predictors for prediction. If both of them have high prediction accuracy and low correlation, they can discriminate between different categories of pixel points. Unlike pairwise PEE, which uses the same predictor for pixel pairs on the diagonal of mutually disjoint $2 \times 2$ pixel blocks. Our proposed algorithm utilizes two distinct predictors for the same pixel to obtain double prediction-errors, and establishes 2D doble prediction-error histogram (2D-DPEH) by counting the frequencies of them. The generated 2D-DPEH can be regarded as consisting of 1D-PEHs. Then for each 1D-PEH, we have to select the expansion bins separately.In \ref{figure:hist-comparison}, our proposed 2D-DPEH distribution, based on double prediction-errors, exhibits sharper feature compared to pairwise PEE on both Lena (plot with smooth gray-scale changes) and Baboon (plot with drastic gray-scale changes). What's more, in order to enhance the speed and performance of proposed method, the selection of expansion bins is determined by dynamic programming (DP) instead of exhaustive search. Finally, as an extension of some existing methos, we confirm the feasibility and superiority of the proposed scheme by combining 2D-DPEH with C-PEE and MHM.

    
    
    

The remaining parts of the paper is as follows. Some PEE-related methods are briefly introduced in \ref{Section II}. Then the details of proposed method is presented in \ref{Section III}, and the experimental results are shown in \ref{Section IV}. Finally, \ref{Section V} gives the conclusion.

\section{Related Work}
\label{Section II}

\subsection{C-PEE} \label{related_work_C-PEE}

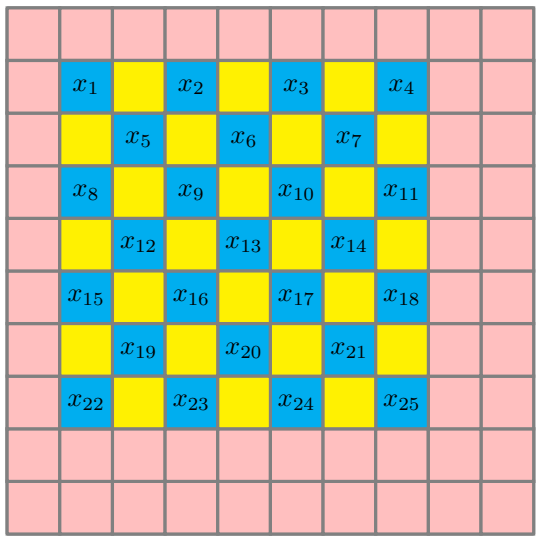
\begin{figure}
    \centering
    \tikzstyle{layer1}=[color = gray, fill = cyan, very thick]
\tikzstyle{layer2}=[color = gray, fill = yellow, very thick]
\tikzstyle{discard}=[color = gray, fill = pink, very thick]

\begin{tikzpicture}[scale = 0.7]
    
    \tikzmath{ 
        function plot_discard(\size) {
            for \x in {0, ..., \size - 1} {
                {
                \draw[discard] (\x, \size - 1) rectangle (\x + 1, \size);
                \draw[discard] (\x, \size - 2) rectangle (\x + 1, \size - 1);
                \draw[discard] (\x, 0) rectangle (\x + 1, 1);
                \draw[discard] (\x, 1) rectangle (\x + 1, 2);
                };
            };
            for \y in {1, ..., \size - 2} {
                {
                \draw[discard] (0, \y) rectangle (1, \y + 1);
                \draw[discard] (1, \y) rectangle (2, \y + 1);
                \draw[discard] (\size - 1, \y) rectangle (\size, \y + 1);
                \draw[discard] (\size - 2, \y) rectangle (\size - 1, \y + 1);
                };
            };
        };
        function plot_layer1(\size, \num) {
            \xs = 2;
            \cnt = 0;
            for \y in {\size - 2, ..., 2} {
                \xs = 3 - \xs;
                for \x in {\xs, \xs + 2, ..., \size - 3} {
                    {
                    \draw[layer1] (\x, \y) rectangle (\x + 1, \y + 1);
                    };
                    \cnt = \cnt + 1;
                    if \cnt <= \num then {
                        int \tmp;
                        \tmp = \cnt;
                        {
                        \node[scale = 1.0] at (\x + 0.5, \y + 0.5) {$x_{\tmp}$};
                        };
                    };
                };
            };
        };
        function plot_layer2(\size) {
            \xs = 1;
            for \y in {\size - 2, ..., 2} {
                \xs = 3 - \xs;
                for \x in {\xs, \xs + 2, ..., \size - 3} {
                     {
                     \draw[layer2] (\x, \y) rectangle (\x + 1, \y + 1);
                     };
                };
            };
        };
        function plot_double_layer(\size, \num){
            {
            \draw[gray, very thick] (0, 0) grid (\size, \size);
            };
            plot_discard(\size);
            plot_layer1(\size, \num);
            plot_layer2(\size);
        };
        plot_double_layer(10, 25);
    };
    
\end{tikzpicture}
    \caption{partition of image pixels}
    \label{fig_double_layer}
\end{figure}

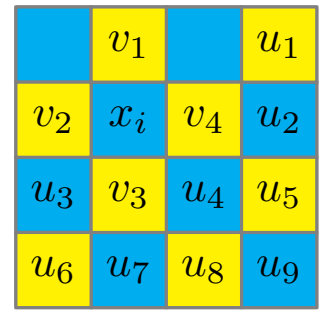
\begin{figure}
    \centering
    \tikzstyle{layer1}=[color = gray, fill = cyan, very thick]
\tikzstyle{layer2}=[color = gray, fill = yellow, very thick]
\tikzstyle{discard}=[color = gray, fill = pink, very thick]

\begin{tikzpicture}
    
    \tikzmath{ 
        function plot_layer1(\size) {
            \xs = 1;
            for \y in {\size - 1, ..., 0} {
                \xs = 1 - \xs;
                for \x in {\xs, \xs + 2, ..., \size - 1} {
                    {
                    \draw[layer1] (\x, \y) rectangle (\x + 1, \y + 1);
                    };
                };
            };
        };
        function plot_layer2(\size) {
            \xs = 0;
            for \y in {\size - 1, ..., 0} {
                \xs = 1 - \xs;
                for \x in {\xs, \xs + 2, ..., \size - 1} {
                     {
                     \draw[layer2] (\x, \y) rectangle (\x + 1, \y + 1);
                     };
                };
            };
        };
        function plot_double_layer(\size){
            {
            \draw[gray, very thick] (0, 0) grid (\size, \size);
            };
            plot_layer1(\size);
            plot_layer2(\size);
            \x = 1.5;
            \y = 2.5;
            {
            \node[scale = 1.7] at (\x, \y) {$x_i$};
            \node[scale = 1.7] at (\x, \y + 1) {$v_1$};
            \node[scale = 1.7] at (\x - 1, \y) {$v_2$};
            \node[scale = 1.7] at (\x, \y - 1) {$v_3$};
            \node[scale = 1.7] at (\x + 1, \y) {$v_4$};
            \node[scale = 1.7] at (\x + 2, \y + 1) {$u_1$};
            \node[scale = 1.7] at (\x + 2, \y) {$u_2$};
            \node[scale = 1.7] at (\x - 1, \y - 1) {$u_3$};
            \node[scale = 1.7] at (\x + 1, \y - 1) {$u_4$};
            \node[scale = 1.7] at (\x + 2, \y - 1) {$u_5$};
            \node[scale = 1.7] at (\x - 1, \y - 2) {$u_6$};
            \node[scale = 1.7] at (\x, \y - 2) {$u_7$};
            \node[scale = 1.7] at (\x + 1, \y - 2) {$u_8$};
            \node[scale = 1.7] at (\x + 2, \y - 2) {$u_9$};
            };
        };
        plot_double_layer(4);
    };
    
\end{tikzpicture}
    \caption{prediction context of $x_i$}
    \label{fig_prediction_context}
\end{figure}

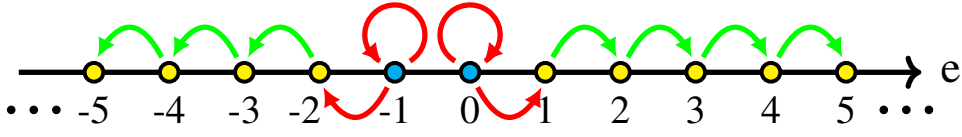
\begin{figure*}
    \centering
    \tikzstyle{expandBin}=[draw, circle, inner sep = 2.5, fill = cyan, line width = 1.5]
\tikzstyle{remainBin}=[draw, circle, inner sep = 2.5, fill = magenta, line width = 1.5]
\tikzstyle{shiftBin}=[draw, circle, inner sep = 2.5, fill = yellow, line width = 1.5]
\tikzstyle{expandArrow}= [draw = red, line width = 2.0, -{Latex[length = 2.5mm, width = 2.5mm]}]
\tikzstyle{shiftArrow} = [draw = green, line width = 2.0, -{Latex[length = 2.5mm, width = 2.5mm]}]
\tikzstyle{loopArrow} = [draw = red, line width = 0.001, -{Latex[length = 2.2mm, width = 2.2mm]}]

\begin{tikzpicture}
    
    \tikzmath{ 
        function plot_edge(\expandFlag, \upFlag, \xStart, \xEnd) {
            if \expandFlag == 1 then {
                \y = 0.2;
                if \upFlag == 0 then {
                    \y = -0.2;
                };
                \xs = 0;
                \xe = 0;
                if \xStart < \xEnd then {
                    \xs = \xStart + 0.1;
                    \xe = \xEnd - 0.05;
                };
                if \xStart > \xEnd then {
                    \xs = \xStart - 0.1;
                    \xe = \xEnd + 0.05;
                };
                \xc = (\xs + \xe) / 2;
                \yc = \y + 0.4;
                if \upFlag == 0 then {
                    \yc = \y - 0.4;
                };
                {
                \draw[expandArrow] (\xs, \y) parabola bend (\xc, \yc) (\xe, \y);
                };
            };
            if \expandFlag == 0 then {
                \y = 0.2;
                if \upFlag == 0 then {
                    \y = -0.2;
                };
                \xs = 0;
                \xe = 0;
                if \xStart < \xEnd then {
                    \xs = \xStart + 0.1;
                    \xe = \xEnd - 0.05;
                };
                if \xStart > \xEnd then {
                    \xs = \xStart - 0.1;
                    \xe = \xEnd + 0.05;
                };
                \xc = (\xs + \xe) / 2;
                \yc = \y + 0.4;
                if \upFlag == 0 then {
                    \yc = \y - 0.4;
                };
                {
                \draw[shiftArrow] (\xs, \y) parabola bend (\xc, \yc) (\xe, \y);
                };
            };
        };
        function plot_loop(\x, \r, \leftFlag) {
            \xx = \x + \r / 2;
            \yy = 0.15;
            {
            \draw[red, line width = 2.0] (\xx, \yy) arc (-60:240:\r);
            };
            \xDes = \xx - \r;
            \yDes = \yy;
            \d = 0.005;
            \xstart = \xDes - \d * 0.642787609687;
            \ystart = \yDes + \d * 0.766044443119;
            if \leftFlag == 0 then {
                \xDes = \xx;
                \yDes = \yy;
                \d = 0.005;
                \xstart = \xDes + \d * 0.642787609687;
                \ystart = \yDes + \d * 0.766044443119;
            };
            {
            \draw[loopArrow] (\xstart, \ystart) -- (\xDes, \yDes);
            };
        };
        function plot_map(\L, \R, \bina, \binb) {
            {
            \draw[->, line width = 2] (\L - 1, 0) -- (\R + 1, 0) node[right, scale = 1.7]{e};
            };
            for \x in {\L, ..., \R} {
                int \tmp;
                \tmp = \x;
                if \x == -2 then {
                    {
                    \node[scale = 1.5] at (\x - 0.2, -0.5){\tmp};
                    };
                };
                if \x != -2 then {
                    {
                    \node[scale = 1.5] at (\x, -0.5){\tmp};
                    };
                };
            };
            for \x in {\L, ..., \R} {
                int \x;
                \x = \x;
                if \x < \bina || \x > \binb then {
                    {
                    \node[shiftBin] at (\x, 0){};
                    };
                };
                if \x == \bina || \x == \binb then {
                    {
                    \node[expandBin] at (\x, 0){};
                    };
                };
                if \x > \bina && \x < \binb then {
                    {
                    \node[remainBin] at (\x, 0){};
                    };
                };
            };
            plot_edge(1, 0, \binb, \binb + 1);
            plot_edge(1, 0, \bina, \bina - 1);
            for \x in {\L, ..., \bina - 2} {
                plot_edge(0, 1, \x + 1, \x);
            };
            for \x in {\binb + 1, ..., \R - 1} {
                plot_edge(0, 1, \x, \x + 1);
            };
            plot_loop(\bina, 0.4, 1);
            plot_loop(\binb, 0.4, 0);
            \xx = 5.5;
            \yy = -0.5;
            {
            \draw[fill = black] (\xx, \yy) circle (0.05);
            \draw[fill = black] (\xx + 0.3, \yy) circle (0.05);
            \draw[fill = black] (\xx + 0.6, \yy) circle (0.05);
            };
            \xx = -5.5;
            {
            \draw[fill = black] (\xx, \yy) circle (0.05);
            \draw[fill = black] (\xx - 0.3, \yy) circle (0.05);
            \draw[fill = black] (\xx - 0.6, \yy) circle (0.05);
            };
        };
        plot_map(-5, 5, -1, 0);
    };
    
\end{tikzpicture}
    \caption{mapping of C-PEE}
    \label{fig_C-PEE_map}
\end{figure*}

The method based on the PEE framework consists of three steps: generating PEH, embedding, extracting and recovering. In C-PEE, a double-layered embedding scheme is adopted, which is to ensure reversibility in the case of utilizing rhombus predictor. This mechanism is to divide the set $S$ consisting of the cover image pixels into two subsets $S_1$ and $S_2$ as shown in Figure \ref{fig_double_layer}, 

where $S_1$ for blue pixels and $S_2$ for yellow ones as $S_1 \cap S_2 = \varnothing$, $S_1 \cup S_2 = S$ and $\forall k \in \{1, 2\}, \forall p_i, p_j \in S_k, i \neq j$, $p_i$ is not adjacent to $p_j$ . First, $S_1$  is used for embedding, and $S_2$ for prediction; then $S_2$ for embedding, $S_1$ for prediction. Each of $S_1$ and $S_2$ takes half of the capacity. While recovering, reversibility can be ensured by first getting $S_2$ according to $S_1$, and then $S_1$ according to $S_2$ . One important thing is that the first two rows and columns and the last row and column in the cover image are not within our consideration. Since the process of twice embedding is similar, we only take the first layer embedding for illustration: we scan the blue pixels from left to right and from top to bottom yields $S_1 = \{x_1, x_2, \cdots , x_N\}$ , and for each pixel $x_i$, the prediction value $\hat x_i$ obtained by using the rhombus predictor is calculated as follows:

\begin{equation}
    \hat x_i = \lceil \frac{v_1 + v_2 + v_3 + v_4}{4} \rceil
    \label{rhomboid_predictor}
\end{equation}
where $v_1, v_2, v_3, v_4$ are the four neighboring yellow pixels of $x_i$ , which is shown in Figure \ref{fig_prediction_context}.
Then the prediction-error $e_i$ is calculated as：
\begin{equation}
    e_i = x_i - \hat x_i
    \label{prediction_error}
\end{equation}
And PEH represented by $h(e)$ is generated by counting the frequency of prediction-errors as: 
\begin{equation}
    h(e) = \#\{1 \leq i \leq N: e_i = e\}, \forall e \in \{-255, \cdots, 255\}
    \label{1D_hist}
\end{equation}
For embedding, C-PEE selects bins -1 and 0 as the expansion bins, and then modifies the prediction-error $e_i$ as:
\begin{equation}
    e_i' = \left\{
    \begin{aligned}
    e_i - 1& , & e_i < -1\\
    e_i - m& , &  e_i = -1\\
    e_i + m& , &  e_i = 0\\
    e_i + 1& , &  e_i > 0\\
    \end{aligned}
    \right.
    \label{C-PEE_error_modify}
\end{equation}
which means $x_i$ is modified as follows：
\begin{equation}
    x_i' = \left\{
    \begin{aligned}
    x_i - 1& , & e_i < -1\\
    x_i - m& , &  e_i = -1\\
    x_i + m& , &  e_i = 0\\
    x_i + 1& , &  e_i > 0\\
    \end{aligned}
    \right.
    \label{C-PEE_xi_modify}
\end{equation}

where $m \in \{0, 1\}$ is a to-be-embedded data, $x_i'$ and $e_i'$ are the modified pixel and prediction-error respectively. The modification of PEH during the embedding is shown in Figure \ref{fig_C-PEE_map}: red arrows are expanded while green arrows are shifted.

As for extracting and recovering，reversibility is guaranteed by inverse scanning. For a markded pixel pixel $x_i'$ , we firstly calculate $e_i'$ and recovery the original prediction-error $e_i$ by $e_i'$ as follows:

\begin{equation}
    e_i = \left\{
    \begin{aligned}
    e_i'& , & e_i' \in \{-1, 0\}\\
    e_i' + 1& , &  e_i' < -1\\
    e_i' - 1& , &  e_i' > 0\\
    \end{aligned}
    \right.
    \label{C-PEE_error_recover}
\end{equation}
then recovery $x_i$ as：
\begin{equation}
    x_i = \left\{
    \begin{aligned}
    x_i'& , & e_i' \in \{-1, 0\}\\
    x_i' + 1& , &  e_i' < -1\\
    x_i' - 1& , &  e_i' > 0\\
    \end{aligned}
    \right.
    \label{C-PEE_xi_recover}
\end{equation}
where the secret message bit can be extracted as $m = 0$ when $e_i \in \{-1, 0\}$, or $m = 1$ when $e_i \in \{-2, 1\}$.

\subsection{MHM}

As a extension of C-PEE，MHM divides pixels into distinct categories by measuring complexity, and adaptively selects expansion bins for each PEH. Specifically, for $x_i$, we take $n_i$ as follows:

\begin{eqnarray}   
    n_i &=& |v_1 - \hat x_i| + |v_2 - \hat x_i| + |v_3 - \hat x_i| + |v_4 - \hat x_i| + \nonumber \\
    &\;& |u_3 - v_3| + |v_3 - u_4| + |u_4 - u_5| + |u_6 - u_7| + \nonumber \\
    &\;& |u_7 - u_8| + |u_8 - u_9| + |v_2 - u_3| + |u_3 - u_6| + \nonumber \\
    &\;& |v_3 - u_7| + |v_4 - u_4| + |u_4 - u_8| + |u_1 - u_2| + \nonumber \\
    &\;& |u_2 - u_5| + |u_5 - u_9| + |v_4 - u_2|
    \label{complexity_calculation}
\end{eqnarray}
where $\{v_1, v_2, v_3, v_4, u_1, u_2, \cdots, u_9\}$ are the pixels in the context of $x_i$ ，which is shown in Figure \ref{fig_prediction_context}. Then $M - 1$ thresholds $\{s_0, s_1, \cdots s_{M - 2}\}$ are determined based on the complexity $n_i$,  the range of $n_i$ is scaled to $M$ values $\{V_0, V_1, \cdots, V_{M - 1}\}$, i.e., $V_0 = [0,s_0], V_1 = [s_0 + 1,s_1], \cdots, V_{M−2} = [s_{M−3} + 1,s_{M−2}], V_{M−1} = [s_{M−2} + 1, \infty)$ .

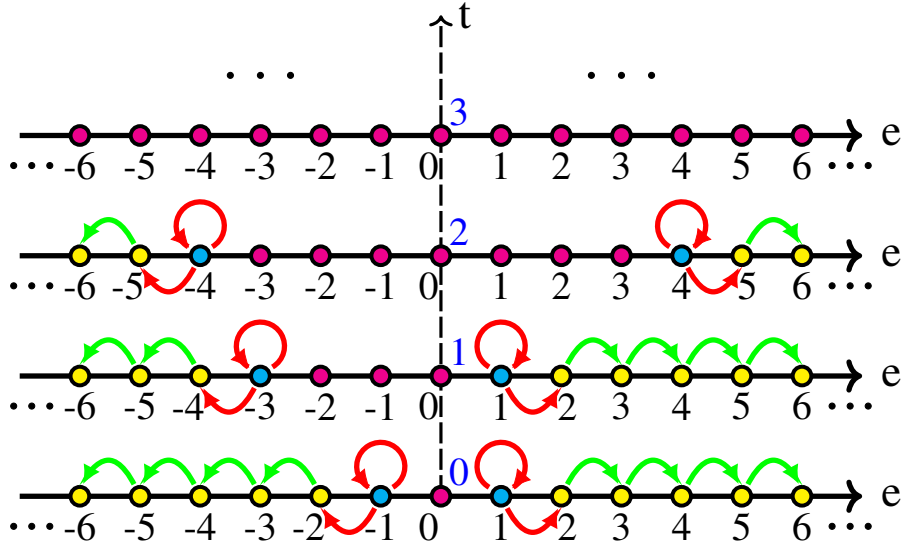
\begin{figure*}
    \centering
    \tikzstyle{expandBin}=[draw, circle, inner sep = 2.5, fill = cyan, line width = 1.5]
\tikzstyle{remainBin}=[draw, circle, inner sep = 2.5, fill = magenta, line width = 1.5]
\tikzstyle{shiftBin}=[draw, circle, inner sep = 2.5, fill = yellow, line width = 1.5]
\tikzstyle{expandArrow}= [draw = red, line width = 2.0, -{Latex[length = 2.5mm, width = 2.5mm]}]
\tikzstyle{shiftArrow} = [draw = green, line width = 2.0, -{Latex[length = 2.5mm, width = 2.5mm]}]
\tikzstyle{loopArrow} = [draw = red, line width = 0.001, -{Latex[length = 2.2mm, width = 2.2mm]}]

\begin{tikzpicture}[scale = 0.8]
    
    \tikzmath{ 
        function plot_edge(\yb, \expandFlag, \upFlag, \xStart, \xEnd) {
            if \expandFlag == 1 then {
                \y = \yb + 0.2;
                if \upFlag == 0 then {
                    \y = \yb - 0.2;
                };
                \xs = 0;
                \xe = 0;
                if \xStart < \xEnd then {
                    \xs = \xStart + 0.1;
                    \xe = \xEnd - 0.05;
                };
                if \xStart > \xEnd then {
                    \xs = \xStart - 0.1;
                    \xe = \xEnd + 0.05;
                };
                \xc = (\xs + \xe) / 2;
                \yc = \y + 0.4;
                if \upFlag == 0 then {
                    \yc = \y - 0.4;
                };
                {
                \draw[expandArrow] (\xs, \y) parabola bend (\xc, \yc) (\xe, \y);
                };
            };
            if \expandFlag == 0 then {
                \y = \yb + 0.2;
                if \upFlag == 0 then {
                    \y = \yb - 0.2;
                };
                \xs = 0;
                \xe = 0;
                if \xStart < \xEnd then {
                    \xs = \xStart + 0.1;
                    \xe = \xEnd - 0.05;
                };
                if \xStart > \xEnd then {
                    \xs = \xStart - 0.1;
                    \xe = \xEnd + 0.05;
                };
                \xc = (\xs + \xe) / 2;
                \yc = \y + 0.4;
                if \upFlag == 0 then {
                    \yc = \y - 0.4;
                };
                {
                \draw[shiftArrow] (\xs, \y) parabola bend (\xc, \yc) (\xe, \y);
                };
            };
        };
        function plot_loop(\yb, \x, \r, \leftFlag) {
            \xx = \x + \r / 2;
            \yy = \yb + 0.15;
            {
            \draw[red, line width = 2.0] (\xx, \yy) arc (-60:240:\r);
            };
            \xDes = \xx - \r;
            \yDes = \yy;
            \d = 0.005;
            \xstart = \xDes - \d * 0.642787609687;
            \ystart = \yDes + \d * 0.766044443119;
            if \leftFlag == 0 then {
                \xDes = \xx;
                \yDes = \yy;
                \d = 0.005;
                \xstart = \xDes + \d * 0.642787609687;
                \ystart = \yDes + \d * 0.766044443119;
            };
            {
            \draw[loopArrow] (\xstart, \ystart) -- (\xDes, \yDes);
            };
        };
        function plot_map(\ys, \t, \L, \R, \bina, \binb, \eqMapFlag) {
            {
            \draw[->, line width = 2] (\L - 1, \ys + \t * 2) -- (\R + 1, \ys + \t * 2) node[right, scale = 1.7]{e};
            };
            for \x in {\L, ..., \R} {
                int \tmp;
                \tmp = \x;
                if \x == \bina - 1 then {
                    {
                    \node[scale = 1.5] at (\x - 0.2, \ys + \t * 2 - 0.5){\tmp};
                    };
                };
                if \x == \binb + 1 then {
                    {
                    \node[scale = 1.5] at (\x + 0.1, \ys + \t * 2 - 0.5){\tmp};
                    };  
                };
                if \x == 0 then {
                    {
                    \node[scale = 1.5] at (\x - 0.2, \ys + \t * 2 - 0.5){\tmp};
                    };  
                };
                if \x != \bina - 1 && \x != \binb + 1 && \x != 0 then {
                    {
                    \node[scale = 1.5] at (\x, \ys + \t * 2 - 0.5){\tmp};
                    };
                };
            };
            for \x in {\L, ..., \R} {
                int \x;
                \x = \x;
                if \x < \bina || \x > \binb then {
                    {
                    \node[shiftBin] at (\x, \ys + \t * 2){};
                    };
                };
                if \x == \bina || \x == \binb then {
                    {
                    \node[expandBin] at (\x, \ys + \t * 2){};
                    };
                };
                if \x > \bina && \x < \binb then {
                    {
                    \node[remainBin] at (\x, \ys + \t * 2){};
                    };
                };
            };
            if \eqMapFlag == 0 then {
                plot_edge(\ys + \t * 2, 1, 0, \binb, \binb + 1);
                plot_edge(\ys + \t * 2, 1, 0, \bina, \bina - 1);
                for \x in {\L, ..., \bina - 2} {
                    plot_edge(\ys + \t * 2, 0, 1, \x + 1, \x);
                };
                for \x in {\binb + 1, ..., \R - 1} {
                    plot_edge(\ys + \t * 2, 0, 1, \x, \x + 1);
                };
                plot_loop(\ys + \t * 2, \bina, 0.4, 1);
                plot_loop(\ys + \t * 2, \binb, 0.4, 0);
            };
            \xx = \R + 0.5;
            \yy = \ys + \t * 2 - 0.5;
            {
            \draw[fill = black] (\xx, \yy) circle (0.05);
            \draw[fill = black] (\xx + 0.3, \yy) circle (0.05);
            \draw[fill = black] (\xx + 0.6, \yy) circle (0.05);
            };
            \xx = \L - 0.5;
            {
            \draw[fill = black] (\xx, \yy) circle (0.05);
            \draw[fill = black] (\xx - 0.3, \yy) circle (0.05);
            \draw[fill = black] (\xx - 0.6, \yy) circle (0.05);
            };
        };
        {
        \draw[->, very thick, dash pattern = on 8pt off 2pt] (0, 0) -- (0, 8) node[right, scale = 1.7]{t};
        };
        plot_map(0, 3, -6, 6, -10, 10, 1);
        plot_map(0, 2, -6, 6, -4, 4, 0);
        plot_map(0, 1, -6, 6, -3, 1, 0);
        plot_map(0, 0, -6, 6, -1, 1, 0);
        for \t in {0, ..., 3} {
            int \t;
            \t = \t;
            {
            \node[scale = 1.5, blue] at (0.3, \t * 2 + 0.4){\t};
            };
        };
        {
        \draw[fill = black] (-3.5, 7.0) circle (0.05);
        \draw[fill = black] (-3, 7.0) circle (0.05);
        \draw[fill = black] (-2.5, 7.0) circle (0.05);
        \draw[fill = black] (3.5, 7.0) circle (0.05);
        \draw[fill = black] (3, 7.0) circle (0.05);
        \draw[fill = black] (2.5, 7.0) circle (0.05);
        };
    };
    
\end{tikzpicture}
    \caption{mapping of MHM}
    \label{fig_MHM_map}
\end{figure*}

Thresholds sequence is determined by computing the $M$-quantile of complexity $n_i$ so that the number of pixels in each category is as uniform as possible, and the formal description of the thresholds calculation is as follows:

\begin{equation}
    s_i = \underset{th}{\arg\min} \ \{\frac{\#\{1 \leq j \leq N: n_j <= th\}}{N} \geq \frac{i + 1}{M}\}
    \label{complexity_thresold}
\end{equation}
Correspondingly, original PEH is split into ${h_0(e), h_1(e), \cdots, h_{M-1}(e)}$ as well, where $h_t(e)$ denotes the frequency of prediction-errors with the local complexities belong to $V_t$:
\begin{equation}
    h_t(e) = \#\{1 \leq i \leq N : e_i = e, n_i = t\}
    \label{MHM_hist}
\end{equation}
Different from C-PEE, which selects fixed expansion bins of 0 and -1, MHM adaptively selects expansion bins for each PEH to reduce distortion. Specifically, the expansion bins are determined through the following optimization problem:
\begin{equation}
    \left\{
    \begin{aligned}
    &minimize \ \frac{\sum_{t = 0} ^ {M - 1}(\sum_{e < bin_a ^ t} h_t(e) + \sum_{e > bin_b ^ t} h_t(e))}{\sum_{t = 0} ^ {M - 1} h_t(bin_a ^ t) + h_t(bin_b ^ t)} \\
    &subject \ to \ \sum_{t = 0} ^ {M - 1} (h_t(bin_a ^ t) + h_t(bin_b ^ t)) \geq EC_{exp} \\
    \end{aligned}
    \right.
    \label{MHM_optimize_problem}
\end{equation}
Here, $bin_{(t, a)}$ and $bin_{(t, b)}$ are the expansion bins selected for the $t$-th PEH, and $EC_{exp}$ is the demand for embedding capacity. Figure \ref{fig_MHM_map} illustrates the modification mechanism of multiple histograms in MHM, where $(bin_{0, a}, bin_{0, b})=(-1, 1)$, $(bin_{1, a}, bin_{1, b})=(-3, 1)$, $(bin_{2, a}, bin_{2, b})=(-4, 4) \cdots (bin_{t, a}, bin_{t, b})=(-\infty, \infty), \forall t > 2$.

\begin{figure}
    \centering
    \includegraphics{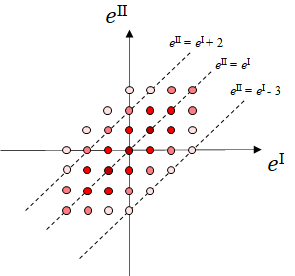}
    \caption{color map of 2D-PEH}
    \label{fig_2D-PEH-colored}
\end{figure}

\section{Proposed Method}
\label{Section III}

In this section, we will describe our proposed method as follows： selection of distinct predictors, 2D-PEH based embedding, determination of optimal parameters and implementation details.

\subsection{Selection of Distinct Predictors}
\label{Section III-A}

We discuss C-PEE firstly, noting the number of pixels with prediction-error $e_i$ as $h(e_i)$ and $(a,b)$ as expansion bins, the embedding capacity and mean square error are:

\begin{equation}
    EC = h(e_a) + h(e_b)
    \label{C-PEE_EC}
\end{equation}
\begin{eqnarray}   
    MSE &=& \frac{1}{n \times m} \times (\sum_{e_i = -255}^{e_a}h(e_i) + \sum_{e_i = e_b}^{255}h(e_i)) \nonumber \\
    &=& 1 - \frac{1}{n \times m} \sum_{e_i = e_a}^{e_b}h(e_i)
    \label{C-PEE_MSE}
\end{eqnarray}

In general, we aim to maximize the embedding capacity $EC$ while minimizing the mean square error $MSE$. Previous works have demonstrated that the higher the accuracy of the predictor, the smaller absolute value of the prediction-error will be, and PEH tends to follow a Laplace distribution, thus the distortion of the embedded image will be smaller. In summary, predictor accuracy is one of the main factors that determine the quality of the embedded image.

After conducting research, we found that all linear predictors, regardless of the accuracy, always perform better in predicting some of the same pixels but not for others. That is to say, most of the linear predictors predicted results with high correlation. However, for pixels with large prediction-errors which are predicted by linear predictors, the nonlinear ones may perform better, while for pixels which are predicted more accurately, the nonlinear ones may not perform well. Therefore, how to apply these two types of predictors to predict these pixels efficiently is of great importance to improve the quality of the embedded image, and we choose both of the linear one and the nonlinear one to calculate the prediction-error simultaneously.






Based on \ref{related_work_C-PEE}, we modify the definition $h$ to:

\begin{equation}
    h(e^{\mathrm{I}},e^{\mathrm{II}}) = \#\{-255 \leq e^{\mathrm{I}},e^{\mathrm{II}} \leq 255:e_i^{\mathrm{I}} = e^{\mathrm{I}},e_i^{\mathrm{II}} = e^{\mathrm{II}}\}
    \label{DP-PEE_hist}
\end{equation}

where $e_i^{\mathrm{I}}$ and $e_i^{\mathrm{II}}$ are prediction-errors derived by the two distinct predictors respectively. $h(e^{\mathrm{I}},e^{\mathrm{II}})$ denotes the frequency of pixels whose double prediction-errors are $e^{\mathrm{I}}$ and $e^{\mathrm{II}}$. As shown in figure 6, the darker the color, the larger the value of histogram $h(e^{\mathrm{I}},e^{\mathrm{II}})$, so as to observe the distribution of $h(e^{\mathrm{I}},e^{\mathrm{II}})$ clearly.

\subsection{2D-PEH Based Embedding}

The 2D-PEH mentioned in \ref{Section III-A} can be regarded as consisting of a number of 1D-PEHs, each of which can be mapped into the function image $e^{\mathrm{II}} = e^{\mathrm{I}} + b, \space b \in [-255,255]$, i.e., the line corresponding to each $b$ can be viewed as one 1D-PEH. Figure \ref{fig_2D-PEH-colored} labels the line $e^{\mathrm{II}} = e^{\mathrm{I}} + b$ when $b = -3, 0, 2$. Our target is to select a pair of expansion bins for each line $e^{\mathrm{II}} = e^{\mathrm{I}} + b$ contained in the 2D-PEH. When reflected on Figure \ref{fig_2D-PEH-colored}, that is, two points $(p_b,q_b)$ and $(r_b, s_b)$ have to be selected for each line $e^{\mathrm{II}} = e^{\mathrm{I}} + b$. It is required that the total distortion is minimized while ensuring that the expansion bins of the 1D-PEHs satisfy the embedding capacity requirement. We denote the embedding as $EC$ and the distortion as $ED$, which can be calculated as follows:

\begin{equation}
    EC = \sum_{b = -255}^{255}(h(p_{b},q_{b}) + h(r_{b},s_{b}))
    \label{DP-PEE_EC}
\end{equation}
\begin{equation}
    ED = \frac{1}{2}EC + \sum_{b=-255}^{255}\sum_{e^{\mathrm{I}}_b < p_b \vee e^{\mathrm{I}}_b > s_b} h(e^{\mathrm{I}}_b,e^{\mathrm{II}}_b) 
    \label{DP-PEE_ED}
\end{equation} 

Then, for $x_i$ , modify its double prediction-errors $(e_i^{\mathrm{I}},e_i^{\mathrm{II}})$ to be $({e_i^{\mathrm{I}}}',{e_i^{\mathrm{II}}}')$ based on $(p_b,q_b)$ and $(r_b, s_b)$ as:
\begin{equation}
    ({e_i^{\mathrm{I}}}',{e_i^{\mathrm{II}}}') = \left\{
    \begin{aligned}
    ({e_i^{\mathrm{I}}},{e_i^{\mathrm{II}}})& , & p_{b} < e_i^{\mathrm{I}} < q_{b}\\
    ({e_i^{\mathrm{I}}} + m,{e_i^{\mathrm{II}}} + m)& , &  e_i^{\mathrm{I}} = q_{b}\\
    ({e_i^{\mathrm{I}}} - m,{e_i^{\mathrm{II}}} - m)& , &  e_i^{\mathrm{I}} = p_{b}\\
    ({e_i^{\mathrm{I}}} + 1,{e_i^{\mathrm{II}}} + 1)& , &  e_i^{\mathrm{I}} > q_{b}\\
    ({e_i^{\mathrm{I}}} - 1,{e_i^{\mathrm{II}}} - 1)& , &  e_i^{\mathrm{I}} < p_{b}\\
    \end{aligned}
    \right.
    \label{DP-PEE_error_modify}
\end{equation}

where $m \in \{0, 1\}$ is to-be-embedded data. Thereafter, $x_i$ is modified to $x_i'$ as follows:

\begin{equation}
    x_i' = \left\{
    \begin{aligned}
    x_i& , & p_{b} < e_i^{\mathrm{I}} < q_{b}  \\
    x_i + m& , &  e_i^{\mathrm{I}} = q_{b}\\
    x_i - m& , &  e_i^{\mathrm{I}} = p_{b}\\
    x_i + 1& , &  e_i^{\mathrm{I}} > q_{b}\\
    x_i - 1& , &  e_i^{\mathrm{I}} < p_{b}\\
    \end{aligned}
    \right.
    \label{DP-PEE_xi_modify}
\end{equation}

\begin{figure*}
    \centering
    \input{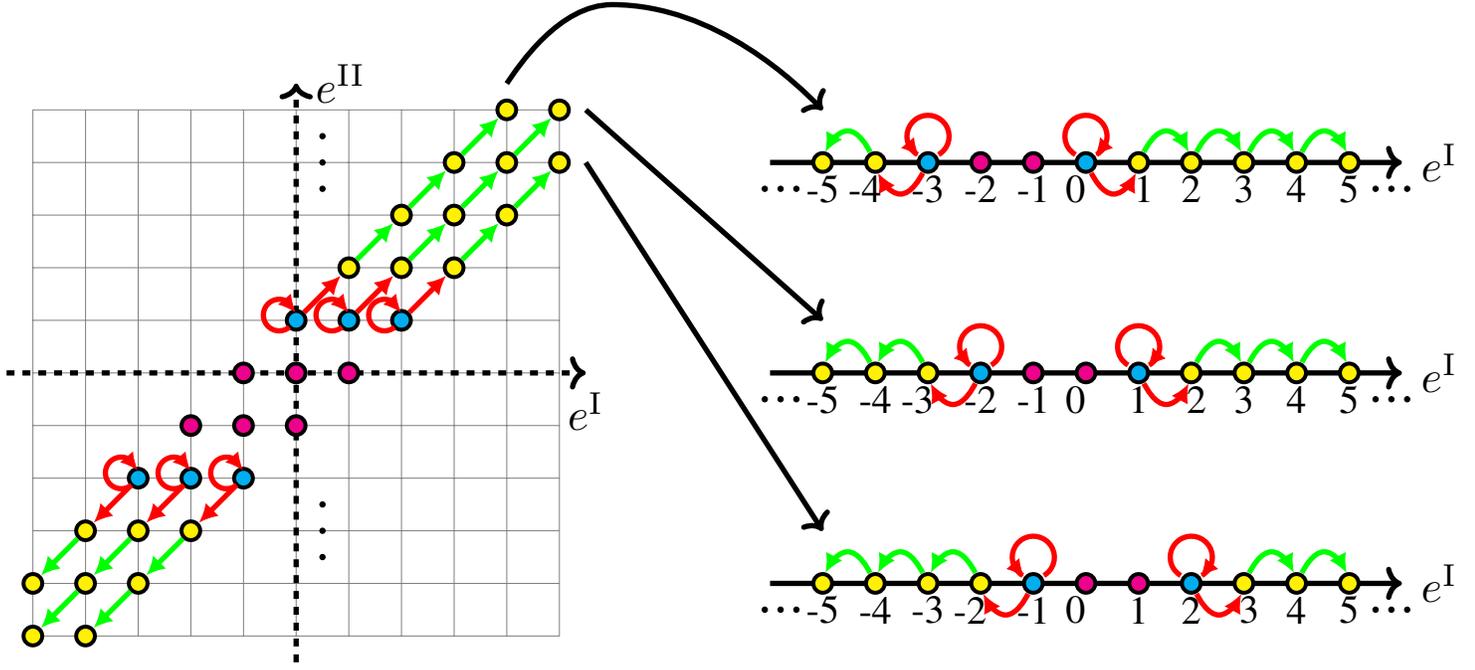}
    \caption{example of 2D-PEH modification}
    \label{fig_2D-PEH-modification}
\end{figure*}

For greater clarity, we show this process in Figure \ref{fig_2D-PEH-modification}.

\begin{figure*}
    \label{C-PEE_vs_proposed}
    \centering
    \begin{minipage}{0.49\linewidth}
	\centering
	\includegraphics[width=0.9\linewidth]{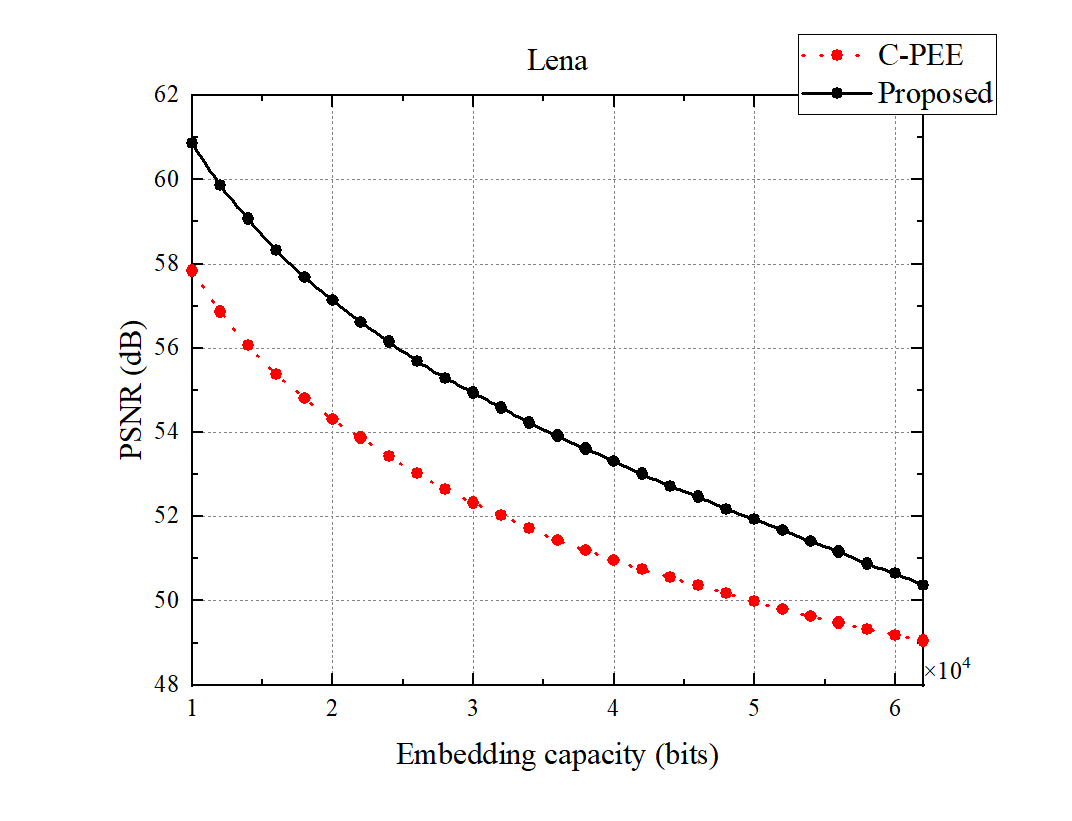}
    \end{minipage}
    \centering
    \begin{minipage}{0.49\linewidth}
	\centering
	\includegraphics[width=0.9\linewidth]{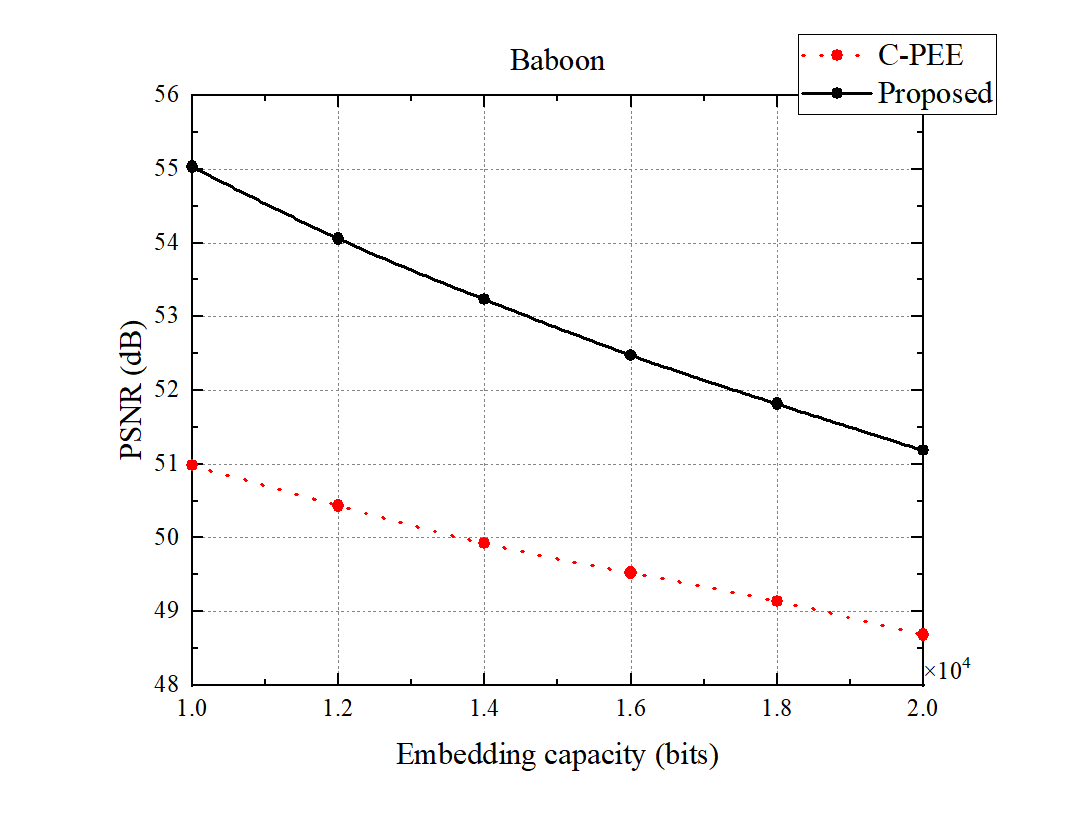}
    \end{minipage}
    \caption{Performance comparison between the proposed method and C-PEE}
\end{figure*}

Due to the use of distinct predictors for prediction, we can classify the pixels in a more detailed way. Therefore, there are more optional expansion bins when embedding, which helps us to increase the capacity and reduce the distortion. To better demonstrate the performance of our proposed method, a comparison of the experimental result with C-PEE on Lena and Baboon is shown in Figure \ref{C-PEE_vs_proposed}, in which C-PEE uses rhombus predictor, double-layered embedding. Our proposed method uses rhombus predictor and MED, double-layered embedding as well. Specifically, $\hat{x_i^{\mathrm{I}}}$ is computed as:

\begin{equation}
\begin{aligned}
\hat{x_i^{\mathrm{I}}} = \lceil \frac{v_1+v_2+v_3+v_4}{4} \rceil, \\
\label{DP-PEE_predictor1}
\end{aligned}
\end{equation}

and $\hat{x_i^{\mathrm{II}}}$ is calculated as：

\begin{equation}
\hat{x_i^{\mathrm{II}}} = \left\{
\begin{aligned}
min(v_3, v_4)& , & u_4 \geq max(v_3, v_4)\\
max(v_3, v_4)& , & u_4 \leq min(v_3, v_4)\\
v_3 + v_4 - u_4&, & otherwise. \\
\end{aligned}
\right.
\end{equation}

The double prediction-errors $(e_i^{\mathrm{I}},e_i^{\mathrm{II}})$ is calculated as:

\begin{equation}
\begin{aligned}(e_i^{\mathrm{I}},e_i^{\mathrm{II}}) = (x_i - \hat{x_i^{\mathrm{I}}}, x_i - \hat{x_i^{\mathrm{II}}})\end{aligned}
\end{equation}

Then, we generate 2D-DPEH as illustrated in \ref{Section III-A} and embed secret message according to \eqref{DP-PEE_error_modify} and \eqref{DP-PEE_xi_modify}. The results in Figure \ref{C-PEE_vs_proposed} has shown that our proposed 2D-DPEH Embedding has a significant improvement in PSNR compared to C-PEE, which is 2.35dB higher on average. The experiments verify the feasibility of the distinct predictors and the superiority of the proposed method.

\begin{figure*}[h]
    \label{flowchart}
    \centering
    \includegraphics[width=0.9\linewidth]{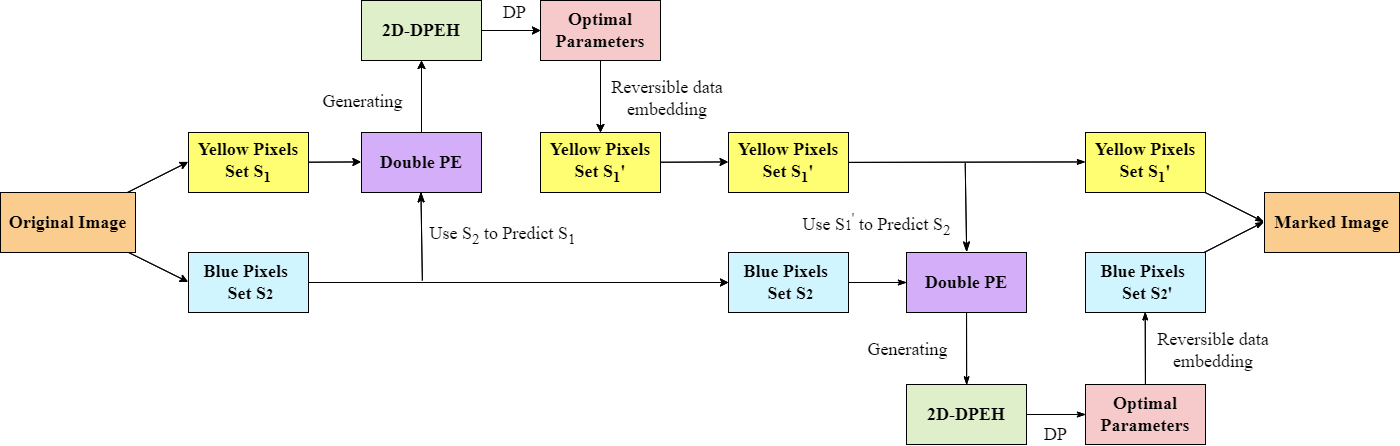}
    \caption{The flow chart of proposed method}
\end{figure*}

Next, we extend the proposed method to multiple 2D-DPEHs by combining it with MHM to further verify the theoretical value of it. The flowchart of the method is shown as \ref{flowchart}. We first calculate the complexity $n_i$ and double prediction-errors $e_i^{\mathrm{I}},e_i^{\mathrm{II}}$ of each pixel $x_i$. Since previous works have demonstrated that the full-enclosing-based predictor performs better than half-enclosing-based predictor in accuracy, the implementation of multiple 2D-PEHs employs two distinct rhombus predictors. For the linear rhombus prediction, we utilize the mean prediction, i.e., $\hat{x_i^{\mathrm{I}}}$ computed by \eqref{DP-PEE_xi_modify}. As for the nonlinear rhombus prediction, we design it based on MED \cite{bib_C-PEE-3}, which first calculates the vertical and horizontal intensities $I_1$ and $I_2$ of the image edges separately:

\begin{equation}
    \left\{
    \begin{aligned}
    I_1 &=& \frac{1}{n \times m} \sum_{i = 0} ^ {n - 1} \sum_{j = 0} ^ {m - 1} (x_{i - 1, j + 1} + x_{i, j + 1} + x_{i + 1, j + 1} \\
    &\ & - x_{i - 1, j - 1} - x_{i, j - 1} - x_{i + 1, j - 1})
    \\
    I_2 &=& \frac{1}{n \times m} \sum_{i = 0} ^ {n - 1} \sum_{j = 0} ^ {m - 1} (x_{i - 1, j - 1} + x_{i - 1, j} + x_{i - 1, j + 1} \\
    &\ & - x_{i + 1, j - 1} - x_{i + 1, j} - x_{i + 1, j + 1})
    \end{aligned}
    \right.
    \label{DP-PEE_predictor2_I}
\end{equation}

The prediction-error $\hat{x_i^{\mathrm{II}}}$ is calculated as follows:

\begin{equation}
    \hat{x_i^{\mathrm{II}}} = \left\{
    \begin{aligned}
    &\lceil \frac{max(v_1,v_2) + min(v_3,v_4)}{2} \rceil ,  min(v_1,v_2) \geq max(v_3,v_4)  \\
    &\lceil \frac{max(v_2,v_3) + min(v_1,v_4)}{2} \rceil ,  min(v_2,v_3) \geq max(v_1,v_4)  \\
    &\lceil \frac{max(v_3,v_4) + min(v_1,v_2)}{2} \rceil ,  min(v_3,v_4) \geq max(v_1,v_2) \\
    &\lceil \frac{max(v_1,v_4) + min(v_2,v_3)}{2} \rceil ,  min(v_1,v_4) \geq max(v_2,v_3)\\
    &\lceil \frac{v_1 + v_3}{2} \rceil ,   min(v_1,v_3) \geq max(v_2,v_4), I_1 \geq I_2\\
    &\lceil \frac{v_2 + v_4}{2} \rceil ,   min(v_2,v_4) \geq max(v_1,v_3), I_1 < I_2\\
    \end{aligned}
    \right.
    \label{DP-PEE_predictor2}
\end{equation}

The 2D-DPEH generated by our proposed method is shown in Figure \ref{figure:hist-comparison} (a) (c). Then the double prediction-errors $e_i^{\mathrm{I}}$ and $e_i^{\mathrm{II}}$ are calculated as follows:

\begin{equation}
    \left\{
    \begin{aligned}
    e_i^{\mathrm{I}} &= x_i - \hat{x_i^{\mathrm{I}}}
    \\
    e_i^{\mathrm{II}} &= x_i - \hat{x_i^{\mathrm{II}}}
    \end{aligned}
    \right.
    \label{DP-PEE_error}
\end{equation}

After that, $M-1$ thresholds are determined based on $n_i$, which divide the thresholds sequence into $M$ intervals $V_0 = [0,s_0], V_1 = [s_0 + 1,s_1], \cdots, V_{M-2} = [s_{M-3} + 1,s_{M-2 }], V_{M-1} = [s_{M-2} + 1, \infty)$ , and $M$ 2D-DPEHs are generated accordingly. Thus the prediction-errors are uniformly partitioned into $M$ 2D-DPEHs:

\begin{equation}
    h_t(e^{\mathrm{I}},e^{\mathrm{II}}) = \#\{-255 \leq e^{\mathrm{I}},e^{\mathrm{II}} \leq 255:e_i^{\mathrm{I}} = e^{\mathrm{I}},e_i^{\mathrm{II}} = e^{\mathrm{II}}, n_i \in V_t\}
    \label{DP-PEE_hist_modifyed}
\end{equation}

\begin{figure*}
    \label{multiple 2D-PEH}
    \centering
    \includegraphics[width=0.9\linewidth]{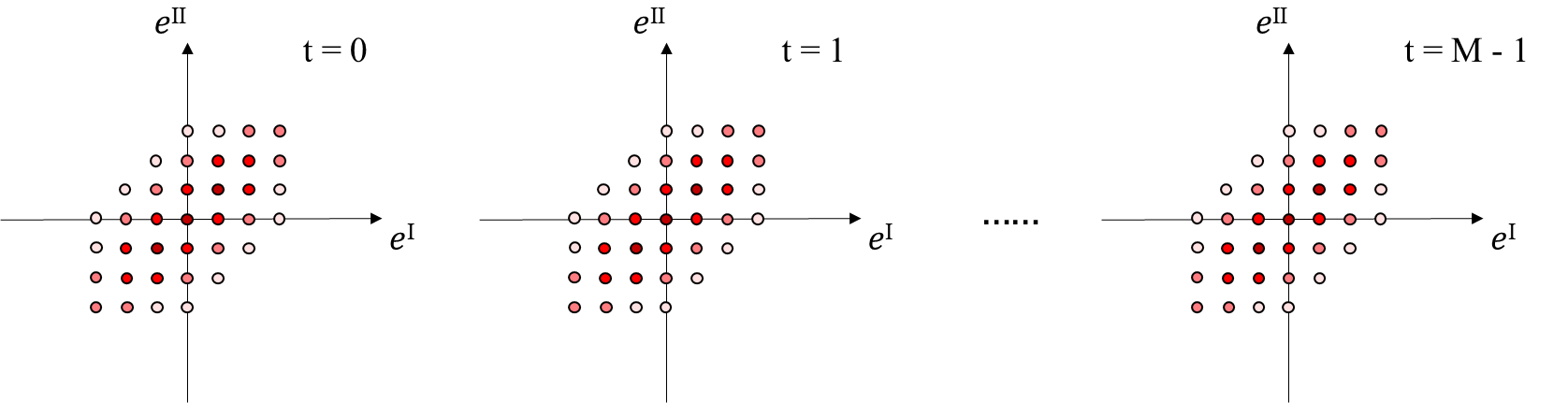}
    \caption{Example of multiple 2D-PEH}
\end{figure*}

At this time, our target is to select a pair of expansion bins for each line $e^{\mathrm{II}} = e^{\mathrm{I}} + b$ contained in every 2D-DPEH, where the selected expansion bins of line $e^{\mathrm{II}} = e^{\mathrm{I}} + b$ belonging to the $t$-th histogram are equivalent to the points $(p_{t,b},q_{t,b})$ and $(r_{t,b}, s_{t,b})$ on the Figure \ref{multiple 2D-PEH}. The total distortion is required to be minimized while ensuring that the expansion bins of the 1D-PEHs corresponding to these points satisfy the capacity demand. The equations for $EC$ and $ED$ are thus modified as follows:

\begin{equation}
    EC = \sum_{t = 0}^{M-1}(\sum_{b = -255}^{255}(h(p_{t,b},q_{t,b}) + h(r_{t,b},s_{t,b}))
    \label{DP-PEE_EC_modifyed}
\end{equation}
\begin{equation}
    ED = \frac{1}{2}EC + \sum_{t = 0}^{M - 1}(\sum_{b=-255}^{255}(\sum_{e^{\mathrm{I}}_{t,b} < p_{t,b} \atop e^{\mathrm{I}}_{t,b} > s_{t,b}} \hspace{-2mm} h(e^{\mathrm{I}}_{t,b},e^{\mathrm{II}}_{t,b}))
    \label{DP-PEE_ED_modifyed}
\end{equation}

Similarly, $({e_i^{\mathrm{I}}},{e_i^{\mathrm{II}}})$ and $x_i$ are recalculated as follows:

\begin{equation}
    ({e_i^{\mathrm{I}}}',{e_i^{\mathrm{II}}}') = \left\{
    \begin{aligned}
    ({e_i^{\mathrm{I}}},{e_i^{\mathrm{II}}})& , & p_{t,b} < e_i^{\mathrm{I}} < q_{t,b}\\
    ({e_i^{\mathrm{I}}} + m,{e_i^{\mathrm{II}}} + m)& , &  e_i^{\mathrm{I}} = q_{t,b}\\
    ({e_i^{\mathrm{I}}} - m,{e_i^{\mathrm{II}}} - m)& , &  e_i^{\mathrm{I}} = p_{t,b}& &n_i \in V_t\\
    ({e_i^{\mathrm{I}}} + 1,{e_i^{\mathrm{II}}} + 1)& , &  e_i^{\mathrm{I}} > q_{t,b}\\
    ({e_i^{\mathrm{I}}} - 1,{e_i^{\mathrm{II}}} - 1)& , &  e_i^{\mathrm{I}} < p_{t,b}\\
    \end{aligned}
    \right.
    \label{DP-PEE_error_modify_modifyed}
\end{equation}
\begin{equation}
    x_i' = \left\{
    \begin{aligned}
    x_i& , & p_{t,b} < e_i^{\mathrm{I}} < q_{t,b}  \\
    x_i + m& , &  e_i^{\mathrm{I}} = q_{t,b}\\
    x_i - m& , &  e_i^{\mathrm{I}} = p_{t,b}& &n_i \in V_t\\\\
    x_i + 1& , &  e_i^{\mathrm{I}} > q_{t,b}\\
    x_i - 1& , &  e_i^{\mathrm{I}} < p_{t,b}\\
    \end{aligned}
    \right.
    \label{DP-PEE_xi_modify_modifyed}
\end{equation}

More specifically, when the value of $x_i$ changes by $1$, the double prediction-errors $e_i^{\mathrm{I}}$ and $e_i^{\mathrm{II}})$ also change by $1$ at the same time, which is reflected in a shift from $(e_i^{\mathrm{I}},e_i^{\mathrm{II}})$ to $(e_i^{\mathrm{I}} \pm 1,e_i^{\mathrm{II}} \pm 1)$ in the 2D-DPEH. Therefore, we can describe this by the following linear function:

\begin{equation}
    e_i^{\mathrm{II}} = e_i^{\mathrm{I}} + b , \ \forall b \in [-255,255]
    \label{DP-PEE_line_eq}
\end{equation}

When we employ only one predictor, all pixels with equal prediction-error have the same decision. Therefore, when the case is two distinct predictors, to ensure consistency before and after decision-making, it is necessary to group all points satisfying $e^{\mathrm{II}} = e^{\mathrm{I}} + b$ into one category and discuss separately. Moreover, when the two predictors are the same, all points on the 2D-DPEH are clustered on the line $e^{\mathrm{II}} = e^{\mathrm{I}}$. In this case, multiple 2D-PEHs are equivalent to $MHM$, in other words, $MHM$ can be regarded as a subset of multiple 2D-DPEHs. Therefore, our proposed method must not be inferior to $MHM$.

Similarly, for multiple 2D-PEHs secret message extraction and image recovery, we first calculate the complexity $n_i$ of the currently modified pixel $\hat{x_i}$, and trace to its corresponding 2D-DPEH $h_t$ based on the thresholds sequence. Then we can get $b$ by following \eqref{DP-PEE_line_eq} to find the corresponding 1D-PEH in $h_t$ and the expansion bins $(p_{t,b}, q_{t,b})$. The original double prediction-errors can be calculated based on $({e_i^{\mathrm{I}}}',{e_i^{\mathrm{II}}}')$ and $(p_{t,b}, q_{t,b})$ as follows:

\vspace{-0.7cm}

\begin{equation}
    ({e_i^{\mathrm{I}}},{e_i^{\mathrm{II}}}) = \left\{
    \begin{aligned}
    &({e_i^{\mathrm{I}}}',{e_i^{\mathrm{II}}}') &p_{t,b} \leq {e_i^{\mathrm{I}}}' \leq q_{t,b}\\
    &({e_i^{\mathrm{I}}}' - 1,{e_i^{\mathrm{II}}}' - 1)&{e_i^{\mathrm{I}}}' > q_{t,b} & & n_i \in V_t\\
    &({e_i^{\mathrm{I}}}' + 1,{e_i^{\mathrm{II}}}' + 1)&{e_i^{\mathrm{I}}}' < p_{t,b}
    \end{aligned}
    \right.
    \label{2D-PEE_error_restore}
\end{equation}

Then, we recover the current pixel as $x_i = \hat{x_i} + e_i^{\mathrm{I}}$. In addition, when ${e_i^{\mathrm{I}}}' \in \{p_{t,b}, q_{t,b}\}$, the embedded data is 0, and when ${e_i^{\mathrm{I}}}' \in \{p_{t,b} - 1, q_{t,b} + 1\}$, the embedded one is 1.

\subsection{Determination of Optimal Parameters}

As mentioned above, for each 1D-PEH in the 2D-PEH, we need to select a pair of expansion bins, which is reflected in Figure \ref{fig_2D-PEH-colored}, i.e., selecting two points with coordinates $(p_{t, b}, q_{t, b})$ and $(r_{t, b}, s_{t ,b})$ respectively. The selection is based on minimizing distortion under the requirement of capacity, which involves solving the following constrained optimization problem:

\begin{equation}
    \left\{
    \begin{aligned}
    &ED = \sum_{t = 0} ^ {M - 1} \sum_{b = -255} ^ {255}(\sum_{e < p_{t, b}} h_t(e, e + b) + \sum_{e > q_{t, b}} h_t(e, e + b)) \\
    &EC = \sum_{t = 0} ^ {M - 1} \sum_{b = -255} ^ {255} h_t(p_{t, b}, p_{t, b} + b) + h_t(q_{t, b}, q_{t, b} + b) \\
    &minimize \ \frac{ED}{EC} \\
    &subject \ to \ EC \geq EC_{exp} \\
    \end{aligned}
    \right.
    \label{MHM_optimize_problem}
\end{equation}

Considering the embedding capacity that most 1D-PEHs can provide is not even sufficient for the bits of auxiliary information needed to record their parameters, lines $e^{\mathrm{II}} = e^{\mathrm{I}} + b$ that correspond to bin heights of 1D-PEHs with a sum of less than $H$ will be categorized as invalid lines which will not be taken into account by us. Here, $H$ is empirically taken as $20$ .

Furthermore, in MHM, parameters are determined by exhaustive search and the time cost is reduced by setting constraints. However, for our proposed algorithm, the time complexity of the exhaustive search is unacceptable, while too many constraints will degrade the quality of the embedded image. Therefore, we utilize DP to solve this problem and optimize the time complexity while achieving better results. In the following part, we will introduce \textbf{Optional Expansion Bins Processing} and \textbf{DP For Optimal Expansion Bins} specifically.

\subsubsection{Optional Expansion Bins Processing}

We first preprocess $EC^{k}_{t,b}$ and ${ED^{k}_{t,b}}$ for all the optional expansion bins of 1D-PEH corresponding to the lines in each 2D-PEH, i.e., $EC^{k}_{t,b}$ and $ED^{k}_{t,b}$ denote the capacity and distortion by selecting the $k$-th optional expansion bins of $e^{\mathrm{II}} = e^{\mathrm{I}} + b$ in $t$-th 2D-PEH, respectively. However, in \ref{Section II}, we have shown that 1D-PEH generated by using the linear rhombus predictor will be much sharper, so that ultimately most of the selections of expansion bins will be relatively concentrated in a small range. Thus there is no need to process out all the optional expansion bins for each line $e^{\mathrm{II}} = e^{\mathrm{I}} + b$, which optimizes the time cost and reduces the amount of auxiliary information. Only bins of $x \in [-14, 14]$ are taken into account.

Specifically, we traverse points of $e_i^{\mathrm{I}} \in [-14,14]$ on each valid line $e^{\mathrm{II}} = e^{\mathrm{I}} + b$ for every 2D-DPEH, which will produce the following three possible options shown in Figure \ref{fig_item _tuple}.

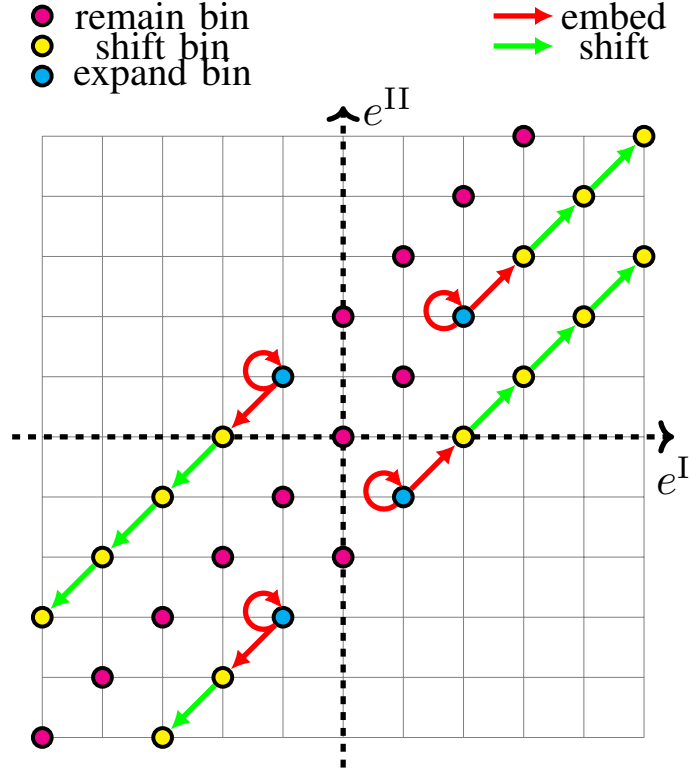
\begin{figure*}
    \centering
    \tikzstyle{expandBin}=[draw, circle, inner sep = 2.5, fill = cyan, line width = 1.5]
\tikzstyle{remainBin}=[draw, circle, inner sep = 2.5, fill = magenta, line width = 1.5]
\tikzstyle{shiftBin}=[draw, circle, inner sep = 2.5, fill = yellow, line width = 1.5]
\tikzstyle{expandArrow}= [draw = red, line width = 2.0, -{Latex[length = 2.5mm, width = 2.5mm]}]
\tikzstyle{shiftArrow} = [draw = green, line width = 2.0, -{Latex[length = 2.5mm, width = 2.5mm]}]
\tikzstyle{loopArrow} = [draw = red, line width = 0.001, -{Latex[length = 2mm, width = 2mm]}]

\begin{tikzpicture}[scale = 0.8]
    \draw[gray, very thin] (-5, -5) grid (5, 5);
    \draw[->, line width = 2, dashed] (-5.5, 0) -- (5.5, 0)  node[below, scale = 1.7]{$e^{\mathrm{I}}$};
    \draw[->, line width = 2, dashed] (0, -5.5) -- (0, 5.5)  node[right, scale = 1.7]{$e^{\mathrm{II}}$};
    \node[expandBin] at (-5, 6) {};
    \node[scale = 1.5] at (-3, 6) {expand bin}; 
    \node[shiftBin] at (-5, 6.5) {};
    \node[scale = 1.5] at (-3, 6.5) {shift bin}; 
    \node[remainBin] at (-5, 7) {};
    \node[scale = 1.5] at (-3, 7) {remain bin}; 
    \draw[expandArrow] (0 + 2.5, 7) -- (1 + 2.5, 7);
    \node[scale = 1.5] at (2 + 2.5, 7) {embed}; 
    \draw[shiftArrow] (0 + 2.5, 6.5) -- (1 + 2.5, 6.5);
    \node[scale = 1.5] at (2 + 2.5, 6.5) {shift}; 
    \tikzmath{ 
        function self_loop(\x, \y, \r){
            \xx = \x - 0.13;
            \yy = \y - 0.13;
            {
            \draw[draw = red, line width = 2.0] (\xx, \yy) arc (310:55:\r);
            };
            \centery = \yy + \r * 0.766044443119;
            \centerx = \xx - \r * 0.642787609687;
            \yend = \centery + \r * 0.258819045103;
            \xend = \centerx + \r * 0.965925826289;
            \d = 0.005;
            \xstart = \xend - \d * 0.642787609687;
            \ystart = \yend + \d * 0.766044443119;
            {
            \draw[loopArrow] (\xstart, \ystart) -- (\xend, \yend);
            };
        };
        function plot(\bina, \binb, \slope){
            int \bina, \binb, \slope;
            \bina = \bina;
            \binb = \binb;
            \slope = \slope;
            for \x in {\bina + 1, ..., \binb - 1} {
                int \x, \y;
                \x = \x;
                \y = \x + \slope;
                if -5 <= \x && \x <= 5 && -5 <= \y && \y <= 5 then {
                    {
                    \node[remainBin] (bin\x\y) at (\x, \y){};
                    };
                };
            };
            if -5 <= \bina - 1 then {
                for \x in {-5, ..., \bina - 1} {
                    int \x, \y;
                    \x = \x;
                    \y = \x + \slope;
                    if \y >= -5 then 
                    {
                        {
                        \node[shiftBin] (bin\x\y) at (\x, \y){};
                        };
                        if \x > -5 && \y > -5 then {
                            int \xlast, \ylast;
                            \xlast = \x - 1;
                            \ylast = \xlast + \slope;
                            {
                            \draw[shiftArrow] (bin\x\y) to (bin\xlast\ylast);
                            };
                        };
                    };
                };
            };
            if \binb + 1 <= 5 then {
                for \x in {\binb + 1, ..., 5} {
                    int \x, \y;
                    \x = \x;
                    \y = \x + \slope;
                    if \y <= 5 then
                    {
                        {
                        \node[shiftBin] (bin\x\y) at (\x, \y){};
                        };
                        if \x > \binb + 1 then {
                            int \xlast, \ylast;
                            \xlast = \x - 1;
                            \ylast = \xlast + \slope;
                            {
                            \draw[shiftArrow] (bin\xlast\ylast) -- (bin\x\y);
                            };
                        };
                    };
                }; 
            };
            int \y, \xnxt, \ynxt;
            if \bina >= -5 then {
                \y = \bina + \slope;
                \xnxt = \bina - 1;
                \ynxt = \xnxt + \slope;
                {
                \node[expandBin] (bin\bina\y) at (\bina, \y){};
                \draw[expandArrow] (bin\bina\y) -- (bin\xnxt\ynxt);
                };
                self_loop(\bina, \y, 0.3);
            };
            if \binb <= 5 then {
                \y = \binb + \slope;
                \xnxt = \binb + 1;
                \ynxt = \xnxt + \slope;
                {
                \node[expandBin] (bin\binb\y) at (\binb, \y){};
                \draw[expandArrow] (bin\binb\y) -- (bin\xnxt\ynxt);
                };
                self_loop(\binb, \y, 0.3);
            };
        };
        plot(-1, 100, 2);
        plot(-100, 2, 0);
        plot(-1, 1, -2);
    }
    
\end{tikzpicture}
    \caption{item illustration}
    \label{fig_item_tuple}
\end{figure*}
\begin{itemize}
    \item Take $(e_i^{\mathrm{I}},e_i^{\mathrm{II}})$ as expansion bin a，$(\infty, \infty)$ as expansion bin b， EC is  $h_t(e_i^{\mathrm{I}},e_i^{\mathrm{II}})$， ED is $\frac{1}{2}h_t(e_i^{\mathrm{I}},e_i^{\mathrm{II}}) + \sum_{e_j^{\mathrm{I}} < e_i^{\mathrm{I}}} h_t(e_j^{\mathrm{I}}, e_j^{\mathrm{II}})$
    \item  Take $(-\infty, -\infty)$ as expansion bin a， $(e_i^{\mathrm{I}},e_i^{\mathrm{II}})$ as expansion bin b，EC is $h_t(x_i,y_i)$ ，ED is $\frac{1}{2}h_t(e_i^{\mathrm{I}},e_i^{\mathrm{II}}) + \sum_{e_j^{\mathrm{I}} > e_i^{\mathrm{I}}} h_t(e_j^{\mathrm{I}},e_j^{\mathrm{II}})$
    \item Take $(e_i^{\mathrm{I}},e_i^{\mathrm{II}})$ as expansion bin a， $(e_j^{\mathrm{I}},e_j^{\mathrm{II}})$ as expansion bin b，$x_i < x_j$ ，EC is $h_t(x_i,y_i) + h_t(x_j,y_j)$ ，ED is $\frac{1}{2}(h_t(e_i^{\mathrm{I}},e_i^{\mathrm{II}}) + h_t(e_j^{\mathrm{I}},e_j^{\mathrm{II}})) + \sum_{e_k^{\mathrm{I}} < e_i^{\mathrm{I}}} h_t(e_k^{\mathrm{I}},e_k^{\mathrm{II}}) + \sum_{e_l^{\mathrm{I}} > e_j^{\mathrm{I}}} h_t(e_l^{\mathrm{I}},e_l^{\mathrm{II}})$
\end{itemize}

For example, now on the line $y = x + 3$ of 2D-DPEH whose complexity belongs to $V_2$, the points $(e^{\mathrm{I}}, e^{\mathrm{II}}))$ of $h_2(e^{\mathrm{I}}, e^{\mathrm{II}}) > 0$ include $\{(-16, -13),(0, 3),(1, 4),(5, 8),(12, 15)\}$:
$h_2(-16,-13) = 150, h_2(0,3) = 100, h_2(1,4) = 300, h_2(5,8) = 200, h_2(12,15) = 250$.

Due to $-16 < -14$, point $(-16, -13)$ is beyond the search range, which is not taken into consideration. Assuming we are currently processing $(1,4)$, as mentioned above, the following cases are possible:

\begin{itemize}
    \item Take $(1,4)$ as expansion bin a，$(\infty, \infty)$ as expansion bin b， EC is $h_2(1,4) = 300$， ED is $\frac{1}{2}h_2(1,4) + h_2(-16, -13) + h_2(0,3) = 400$
    \item Take $(-\infty, -\infty)$ as expansion bin a， $(1,4)$ as expansion bin b， EC is $h_2(1,4) = 300$， ED is $\frac{1}{2}h_2(1,4) + h_2(5, 8) + h_2(12,15) = 600$
    \item Take $(1,4)$ as expansion bin a， $(5,8)$ as expansion bin b， EC is $h_2(1,4) + h_2(5,8) = 500$ ， ED is $\frac{1}{2}(h_2(1,4) + h_2(5,8)) + h_2(-16, -13) + h_2(0,3) + h_2(12,15) = 750$
    \item Take $(1,4)$ as expansion bin a， $(12,15)$ as expansion bin b ， EC is $h_2(1,4) + h_2(12,15) = 550$ ， ED is $\frac{1}{2}(h_2(1,4) + h_2(12,15)) + h_2(-16, -13) + h_2(0,3) = 525$
\end{itemize}

For $(0,3),(5,8),(12,15)$, similarly, we add the above cases to the decision-making alternatives of this group for the upcoming DP.

\subsubsection{DP For Optimal Expansion Bins}

To delineate more clearly, the parameters of DP For Optimal Expansion Bins are defined using the symbols shown in Table \ref{table:symbol}.

\begin{table*}
    \caption{definition of symbols in DP}
    \centering
    \begin{tabular}{cc}
    \hline
    symbol  & definition \\
    \hline
    $f_{t, i, j}$  & \makecell[c]{the minimum distortion caused by the first $t$ 2D-PEHs, the $t$-th one considers the lines with intercepts of $-255 \sim i$, \\ while satisfying the embedding capacity requirement of $j$.}\\
    $trans_{t, i, j}$  & the transfer source of $f_{t, i, j}$  \\
    $exBin_{t, i, k}$ & the $k$-th optional expansion bins of a line with intercept $i$ in the $t$-th 2D-DPEH. \\
    $EC_{t, i, k}$ & the embedding capacity offered by $exBin_{t, i, k}$.\\
    $ED_{t, i, k}$ & the distortion caused by $exBin_{t, i, k}$.\\
    $n_{t, i}$ & the number of optimal expansion of the line with intercept $i$ in the $t$-th 2D-DPEH.  \\
    $optBin_{t, i}$ & the optimal expansion of a line with intercept $i$ in the $t$-th 2D-DPEH. \\
    $\triangle$ & a increment to prevent the array out of bounds. \\
    \hline
    \end{tabular}
    \label{table:symbol}
\end{table*}

We first design the state and boundary conditions, and let $f_{t, i, j}$ denote the minimum distortion caused by the first $t$ 2D-DPEHs, the $t$-th one only consider lines with $-255 \sim i$ intercepts, and based on the condition of the embedding capacity requirement is $j$. Since the sum of the capacity provided by the expansion bins is allowed to exceed $EC_{exp}$, the maximum value $j$ of the DP array $f_{t, i, j}$ is taken as $EC_{exp} + \triangle$ in the design of the state boundary, where $\triangle$ is a increment to prevent the sum from being out-of-bounds, $\triangle$ is taken as $2,000$ based on several experiments. Obviously, if no expansion bins are selected, the total EC and ED are both 0, so the boundary condition is as follows:

\begin{equation}
    f_{-1, 255, j} = \left\{
    \begin{aligned}
    0 & , & j = 0 \\
    \infty & , &  j \neq 0
    \end{aligned}
    \right.
    \label{DP-PEE_DP_cd}
\end{equation}

Then the state-transition equation is designed as follows:

\begin{equation}
    f_{t, i, j} = \left\{
    \begin{aligned}
    \underset {k}{min}(f_{t, i-1, j}, f_{t, i-1, j - EC_{t, i, k}} + ED_{t, i, k})& , & i > -255\\
    \underset {k}{min}(f_{t - 1, 255, j}, f_{t - 1, 255, j - EC_{t, i, k}} + ED_{t, i, k})& , &  i = -255
    \end{aligned}
    \right.
    \label{DP_trans_equation}
\end{equation}

Based on the boundary conditions \ref{DP-PEE_DP_cd}, iterate continuously according to the state-transition equation \ref{DP_trans_equation}, and end up getting $f_{M - 1, 255, j}$, which denotes the minimum distortion under the condition that the embedding capacity requirement is $j$, the specific process of forward state-transition is shown as Algorithm \ref{algorithm:DP-transition}.

Eventually, we will backtrack to obtain the optimal parameters, in addition to determining the minimum distortion, the exact indexes of selected expansion bins for embedding are desired as well. So it is necessary to record the transfer paths during the DP process, let $trans_{t, i, j}$ denote the transfer source of $f_{t, i, j}$:

\begin{equation}
    trans_{t, i, j} = \left\{
    \begin{aligned}
    \underset {k}{argmin}\{f_{t, i - 1, j - EC_{t, i, k}} + ED_{t, i, k}\}& , & i > -255\\
    \underset {k}{argmin}\{f_{t - 1, 255, j - EC_{t, i, k}} + ED_{t, i, k}\}& , &  i = -255
    \end{aligned}
    \right.
    \label{DP-PEE_DP_trans}
\end{equation}

Once the DP is finished, we determine the optimal embedding capacity (the one that satisfies the capacity requirements and minimizes distortion) that the expansion bins are able to offer by the following equation, denoted as $EC^{\star}$:

\begin{equation}
    EC ^ {\star} = \underset {j \geq EC_{exp}}{\arg\min}\{f_{M - 1, 255, j}\}
    \label{DP-PEE_DP_EC_star}
\end{equation}

Then we obtain the above-mentioned indexes by reverse-traversing the array $trans$, the details are shown in Algorithm \ref{algorithm:DP-reverse}.

Considering that DP needs a huge space to store the state information, we decide to optimize the space complexity by scrolling array. DP topic is a bottom-up expansion technique,  according to \eqref{DP_trans_equation},  we find that $f_{t, i, j}$ is transferred from only $f_{t, i - 1, k}$ or $f_{t - 1, 255, k}$. In other words, the state information stored more previously is useless for $f_{t, i, j}$, which means that some parts of the stored state information can be shed as DP progresses. More specifically, we redefine the states $f_j, g_j$ as follows:

\begin{equation}
    \left\{
    \begin{aligned}
    g_j &= {\min}\{f_j, \underset {k}{\min}\{f_{j - EC_{t, i, k}} + ED_{t, i, k}\}\}
    \\ 
    f_j &= g_j
    \end{aligned}
    \right.
    \label{DP-PEE_DP_rolling_array}
\end{equation}

where $f_j$ and $g_j$ represent the dp result of the last group and the dp result of the current one, $j$ denotes the size of the capacity and $k$ denotes the $k$-th pair of optional expansion bins that is being processed in the current group. The meaning of Equation \eqref{DP-PEE_DP_rolling_array} is that $g_j$ is obtained by transforming from $f_j$ first, and then $g_j$ is assigned to $f_j$ to prepare for the transfer of the next round, the process will continue until getting all of the optimal expansion bins. It can be shown that the space complexity of the proposed method is reduced from the initial $O(M \times 511 \times EC_{exp})$ to $O(EC_{exp})$ ($EC_{exp}$ denotes the expected embedding capacity), thus saving a lot of space.

\begin{algorithm}
\caption{forward state-transition}
\label{algorithm:DP-transition}
\begin{algorithmic}[1] 
\REQUIRE ~~\\ 
    $n_{t, i}$ , $EC_{exp}$, $\triangle$, $EC_{t, i, k}$, $ED_{t, i, k}$. \\
\ENSURE ~~\\ 
    $f_{t, i, j}$;
    $trans_{t, i, j}$;
\STATE initialize $f_{0, 0} \leftarrow 0$;
\FOR{$j \leftarrow 0$ to $EC_{exp} + \triangle$}
    \STATE initialize $f_{-1, 255, j} \leftarrow \infty$;
\ENDFOR
\FOR{$t \leftarrow 0$ to $M - 1$}
    \FOR{$i \leftarrow -255$ to $255$}
        \STATE $t' \leftarrow t$;
        \STATE $i' \leftarrow i - 1$;
        \IF{$i' \leftarrow -256$}
            \STATE $t' \leftarrow t' - 1$;
            \STATE $i' \leftarrow 255$;
        \ENDIF
        \FOR{$j \leftarrow 0$ to $EC_{exp} + \triangle$}
            \STATE $f_{t, i, j} \leftarrow f_{t', i', j}$;
            \STATE $trans_{t, i, j} \leftarrow -1$;
            \FOR{$k \leftarrow 0$ to $n_{t, i} - 1$}
                \IF{($j \geq EC_{t, i, k}$) and ($f_{t', i', j - EC_{t, i, k}} + ED_{t, i, k} < f_{t, i, j}$)}
                    \STATE $f_{t, i, j} \leftarrow f_{t', i', j - EC_{t, i, k}} + ED_{t, i, k}$;
                    \STATE $trans_{t, i, j} \leftarrow k$;
                \ENDIF
            \ENDFOR
        \ENDFOR
    \ENDFOR
\ENDFOR
\RETURN $f_{t, i, j}$, $trans_{t, i, j}$; 
\end{algorithmic}
\end{algorithm}

\begin{algorithm}
\caption{back state-transition}
\label{algorithm:DP-reverse}
\begin{algorithmic}[1] 
\REQUIRE ~~\\ 
    $f_{t, i, j}$, $trans_{t, i, j}$, $EC_{t, i, k}$, $exBin_{t, i, k}$;\\
\ENSURE ~~\\ 
    $optBin_{t, i}$;\\
\STATE initialize $j \leftarrow \underset{p}{\arg\min}\{f_{M - 1, 255, p}\}$;
\FOR{$t \leftarrow M - 1$ to $0$}
    \FOR{$i \leftarrow 255$ to $-255$}
        \IF{$trans_{t, i, j} \neq -1$}
            \STATE $optBin_{t, i} \leftarrow exBin_{t, i, trans_{t, i, j}}$;
            \STATE $j \leftarrow j - EC_{t, i, trans_{t, i, j}}$;
        \ELSE
            \STATE $optBin_{t, i} \leftarrow (-\infty, \infty)$;
        \ENDIF
    \ENDFOR
\ENDFOR
\RETURN $optBin_{t, i}$; 
\end{algorithmic}
\end{algorithm}

Now assume that the $b$-th line in the $t$-th 2D-DPEH contains $c_{t,b}$ pairs of optional expansion bins. Compared to the exhaustive search, DP optimizes the time cost from $ \prod_{t=0}^{M-1} \prod_{b = -255}^{255} c_{t,b}$ to $ \sum_{t = 0}^{M-1} \sum_{b = -255}^{255} c_{t,b} \times EC$. In short, the time complexity is now optimized by us from exponential to polynomial level. When in the embedding capacities of 10,000 bits on image Lena, the exhaustive search theoretically takes more than $10^{7000}s$, but the experimental results show that the proposed method optimized by DP only takes about $30s$ on C++, which definitely proves the effectiveness of DP.

\subsection{Data Embedding}

After limiting the search range of the expansion bin and excluding invalid lines, the necessary auxiliary information embedded in each layer embedding includes:
\begin{itemize}
    \item the flag bits: to help the receiver distinguish between invalid and valid lines, '$0$' is invalid and '$1$' is valid line,
    \item the parameters: the expansion bins of valid line correspond to the abscissa $e^{\mathrm{I}}$ in \ref{fig_2D-PEH-colored}
    \item the complexity thresholds sequence: to divide multiple 2D-DPEHs
    \item the index $N_{end}$
    \item the parameter $S_{CLM}$
    \item the compressed location map CLM (SCLM bits): to record the overflow or underflow pixels
\end{itemize}

\begin{figure}
\begin{center}
\begin{tabular}{c}
 $\cdots \cdots$ \\
 0000000000000000000000000000000\\
 0000000000000000000000000000000\\
 0000000000000000000000000000000\\
 0000000000000000000000000000000\\
 0000000000000000111110000000000\\
 0000000000000000000000000000000\\
 0000000000000000000000000000000\\
 0000000000000000000000000000000\\
 $\cdots \cdots$
\end{tabular}
\end{center}
\caption{Example of flag bits information} 
\label{continuous bits}
\end{figure}

where the flag bits accounts for the vast majority of the auxiliary information. In order to make the total number of auxiliary information bits fall within an acceptable range, we choose to compress the auxiliary information. By analyzing 2D-DPEHs, we find that valid lines cluster together in each 2D-DPEH. Reflected in the flag bits is a high number of sequential $0$(or $1$), as shown in Figure \ref{continuous bits}, which means that some compression algorithms such as RLE and LZMA can achieve better compression ratio. At the same time, the experimental results also prove that the final auxiliary information bits can be maintained at a lower level (less than HMM-PRH) by LZMA in Lena, $10,000$ capacity.

Next, we take the first layer of embedding as an example to describe the embedding of the proposed method: First, calculate $(e_1^{\mathrm{I}},e_2^{\mathrm{I}} \cdots e_n^{\mathrm{I}})$， $(e_1^{\mathrm{II}},e_2^{\mathrm{II}} \cdots e_n^{\mathrm{II}})$， $(n_1,n_2 \cdots n_N)$, $(s_0, s_1, \cdots, s_{M - 2})$
and thus obtain
$\{h_t\ | 0 \leq t \leq M-1\}$，
$\{(A_{t,b}, B_{t,b}), (C_{t,b}, D_{t,b}) \ | \ 0 \leq t \leq M- 1, -255 \leq b \leq 255, e^{\mathrm{II}} = e^{\mathrm{I}} + b\}$ and auxiliary information.

The method described above is used to perform embedding. The secret message is embedded into $(x_1, \cdots ,x_{N'})$, and the auxiliary information of twice embedding is stored in last two rows. Let $S_{LSB}$ be the $LSBs$ of the initial $S_{aux}$ pixels in the first row, then continue to embed $S_{LSB}$ starting from $x_{N' + 1}$. Here, $x_{N_{end}}$
is the last processed pixel. Finally, the marked image is obtained by replacing LSBs of the first $S_{aux}$ pixels in the last row with auxiliary information. To facilitate data extraction and image restoration, we agree to store $S_{aux}$ in the first 10 bits of the auxiliary information.

\subsection{Data Extraction and Image Restoration}

Still taking the first layer of embedding as an example, at the beginning, we obtain $S_{aux}$ by taking LSB of the first $10$ pixels in the last row, and then we take LSB of $11$ to $S_{aux}$ pixels and decompress them to determine the auxiliary information. After that, the $S_{LSB}$ of $(x_{N' + 1}, \cdots, x_{N_{end}})$ is extracted and restored by scanning in reverse order based on above-mentioned method, and we realize restoration for the last row by replacing LSBs of first $S_{aux}$ pixels with extracted $S_{LSB}$. Finally, also in reverse
scanning order, extract the embedded secret message from $(x_1, \cdots, x_{N'})$ and meanwhile restore these pixels according to \ref{Section III}.
It is important to note that for each overflow or underflow pixel $x_i$ recorded in $CLM$,  modify its value to $255$ if $x_i = 254$, or $0$ if $x_i = 1$.

\begin{figure*}
    \label{MHM_vs_proposed}
    \centering
    \begin{minipage}{0.49\linewidth}
	\centering
	\includegraphics[width=0.9\linewidth]{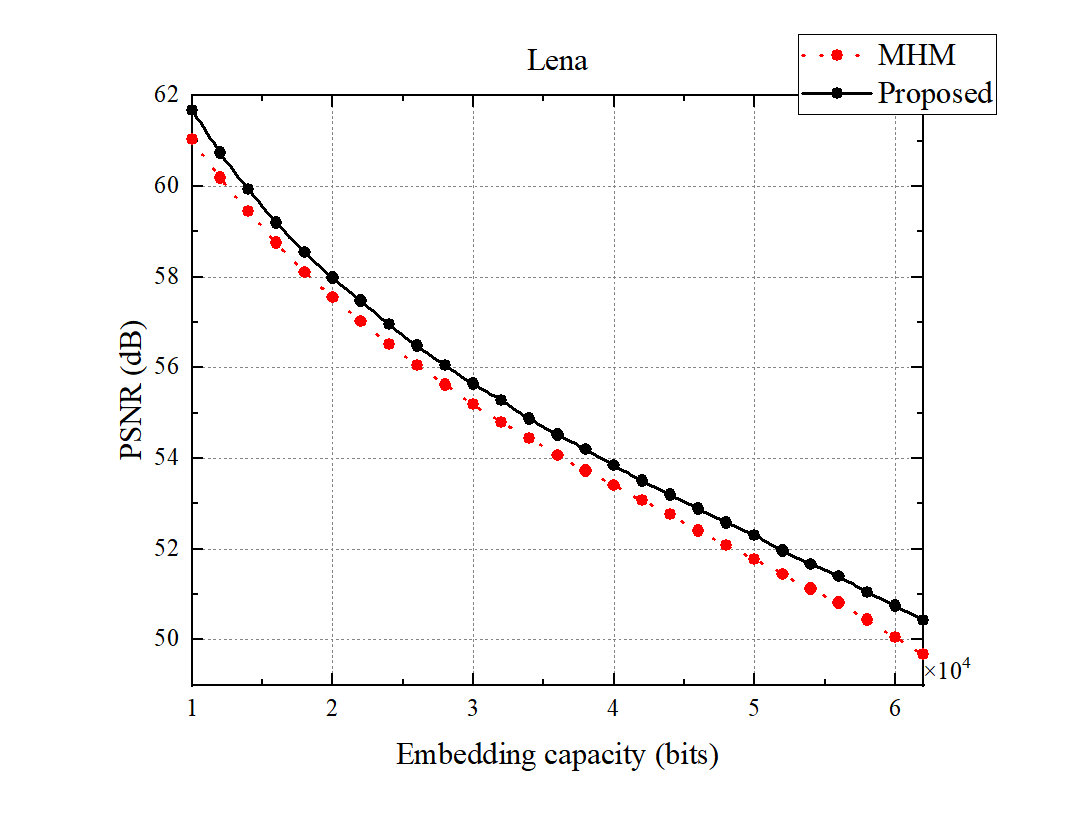}
    \end{minipage}
    \centering
    \begin{minipage}{0.49\linewidth}
	\centering
	\includegraphics[width=0.9\linewidth]{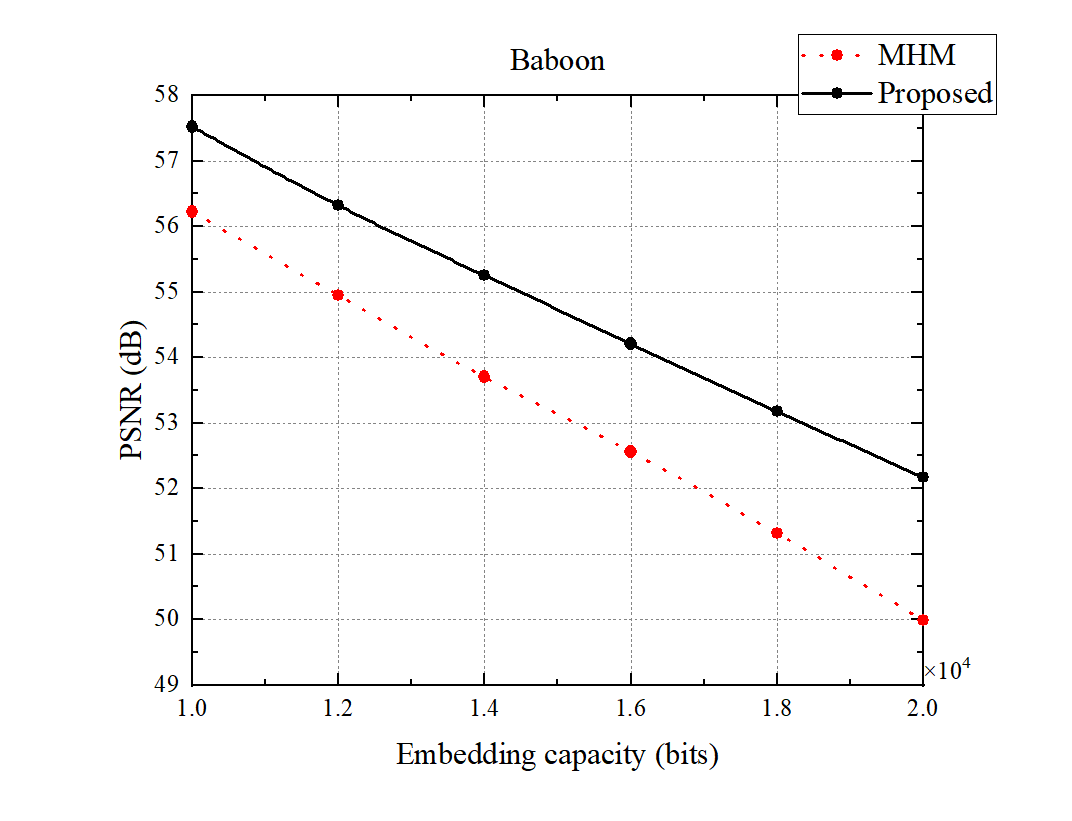}
    \end{minipage}
    \caption{Performance comparison between the proposed method and MHM}
\end{figure*}

From Figure \ref{MHM_vs_proposed}, even though the embedded image quality has achieved a very high level of performance by adopting MHM, the proposed method still improves MHM with an average PSNR gain of 0.50dB and 1.65dB, respectively. The experiment results prove the feasibility of extending 2D-DPEH to multiple 2D-DPEHs and the superiority of the proposed method.

\section{EXPERIMENTAL RESULTS}
\label{Section IV}

\begin{table*}
    \caption{Comparison of PSNR (in dB) between the Proposed Method and the Original MHM [x], MHM-PRE [y], in an Embedding Capacity of 10,000 bits}
    \label{tabel_PSNR_1W}
    \begin{center}
    \begin{tabular}{cccccccc} 
    \hline
    image & C-PEE & Pairwise PEE & MHM & High capacity MHM & Optimal MHM &  MHM-PRE & Proposed \\
    \hline
    Lena & 57.84 & 59.75 & 61.03 & 61.01 & 61.04 & 61.33 & 61.68\\ 
    Baboon & 50.98 & 55.21 & 56.22 & 56.23 & 56.25 & 56.90 & 57.52\\ 
    Barbara & 55.98 & 59.48 & 61.41 & 61.36 & 61.67 & 61.83 & 62.10 \\
    Boat & 53.98 & 57.55 & 58.62 & 58.65 & 58.97 & 59.10 & 60.16\\
    Airplane & 63.9 & 63.76 & 63.87 & 63.89 & 63.90 & 64.26 & 64.25\\ 
    Peppers & 53.99 & 56.21 & 59.06 & 59.07 & 59.37 & 59.60 & 60.22 \\
    \hline
    Average & 56.11 & 58.66 & 60.04 & 60.04 & 60.20 & 60.50 & 60.99 \\ 
    \hline
    
    \end{tabular}
  \end{center}
\end{table*}

\begin{table*}
    \caption{Comparison of PSNR (in dB) between the Proposed Method and the Original MHM [x], MHM-PRE [y], in an Embedding Capacity of 20,000 bits}
    \label{tabel_PSNR_2W}
    \begin{center}
    \begin{tabular}{cccccccc} 
    \hline
     image & C-PEE & Pairwise PEE & MHM & High capacity MHM & Optimal MHM &  MHM-PRE & Proposed \\
    \hline
    Lena & 54.31 & 56.21 & 57.56 & 57.55 & 57.64 & 57.79 & 57.98\\ 
    Baboon & 48.68 & 49.89 & 49.99 & 50.96 & 50.41 & 50.89 & 52.17\\ 
    Barbara & 53.52 & 56.22 & 57.67 & 57.68 & 57.79 & 57.96 & 58.25 \\
    Boat & 50.74 & 53.32 & 54.58 & 54.76 & 54.85 & 55.03 & 55.96\\
    Airplane & 55.85 & 60.15 & 60.55 & 60.49 & 60.60 & 60.75 & 60.75\\ 
    Peppers & 50.74 & 52.83 & 55.12 & 55.25 & 55.37 & 55.55 & 56.12 \\
    \hline
    Average & 52.31 & 54.77 & 55.91 & 56.12 & 56.11 & 56.33 & 56.87\\
    \hline
    \end{tabular}
  \end{center}
\end{table*}

In this section, we mainly evaluate the performance of the proposed method by comparing it method with C-PEE, Pairwise PEE, original MHM, and several state-of-the-art works based on MHM, all of which adopt rhombus predictor, double-layered embedding. The M of MHM methods are taken as 16. Six standard 512 × 512 sized gray-scale images including Lena, Baboon, Barbara, Boat, Airplane and Peppers downloaded from the USC-SIPI image database are selected for experiments. At the same time, experimental results and parameters of $10,000$ and $20,000$ capacity are listed in the Table \ref{tabel_reoccur}. Owing to space constraints, we only list the parameters $(p_{t,b},r_{t,b})$ when $(0 \leq t \leq 2, -2 \leq b \leq 2)$. Here, $(p_{t,b},r_{t,b})$ has the same meaning as in \ref{Section III}.

\begin{table}
    \caption{The determined parameters $p_{t,b}$ , $r_{t,b}(0 \leq t \leq 2, -2 \leq b \leq 2)$ of our method. For the blue (first layer embedding) and yellow (second layer embedding) pixels of image Lena with 10,000 and 20,000 Embedding Capacities}
    \label{tabel_reoccur}
    \begin{center}
    \begin{tabular}{c|c|c|c|c} 
    \hline
    \hline
     \textbf{} & \multicolumn{2}{c|}{\textbf{EC = 10,000 bits}}  & \multicolumn{2}{c}{\textbf{EC = 20,000 bits}}\\
       & \multicolumn{2}{c|}{blue \quad \; yellow}  & \multicolumn{2}{c}{blue \quad \; yellow}\\
    \hline
     \hline
    $(p_{0,-2}, r_{0,-2})$ & \multicolumn{2}{c|}{(-4, 6) \quad (-4, 3)} & \multicolumn{2}{c}{(-4, 2) \quad (-2, 1)}\\ 
    \hline
    $(p_{0,-1}, r_{0,-1})$ & \multicolumn{2}{c|}{(-2, 3) \quad (-4, 3)} & \multicolumn{2}{c}{(-2, 1) \quad (-2, 1)}\\ 
    \hline
    $(p_{0,0}, r_{0,0})$ & \multicolumn{2}{c|}{(-3, 2) \quad (-2, 2)} & \multicolumn{2}{c}{(-1, 1) \quad (-2, 2)}\\ 
    \hline
    $(p_{0,1}, r_{0,1})$ & \multicolumn{2}{c|}{(-2, 1) \quad (-3, 1)} & \multicolumn{2}{c}{(-1, 1) \quad (-2, 1)}\\ 
    \hline
    $(p_{0,2}, r_{0,2})$ & \multicolumn{2}{c|}{(-2, 2) \quad (-2, 3)} & \multicolumn{2}{c}{(-2, 2) \quad (-2, 1)}\\
    \hline
    \hline
    $(p_{1,-2}, r_{1,-2})$ & 
     \multicolumn{2}{c|}{(-2, 6) \quad (-3, 2)} & 
     \multicolumn{2}{c}{(-2, 2) \quad (-2, 2)}\\
     \hline
    $(p_{1,-1}, r_{1,-1})$ & 
     \multicolumn{2}{c|}{(-3, 2) \quad (-4, 3)} & 
     \multicolumn{2}{c}{(-3, 2) \quad (-3, 2)}\\
     \hline
    $(p_{1,0}, r_{1,0})$ & 
     \multicolumn{2}{c|}{(-5, 3) \quad (-4, 4)} & 
     \multicolumn{2}{c}{(-3, 2) \quad (-4, 2)} \\
     \hline
    $(p_{1,1}, r_{1,1})$ & 
     \multicolumn{2}{c|}{(-4, 4) \quad (-4, 5)} & 
     \multicolumn{2}{c}{(-4, 2) \quad (-4, 2)} \\
     \hline
    $(p_{1,2}, r_{1,2})$ & 
     \multicolumn{2}{c|}{(-3, 5) \quad (-3, 3)} & 
     \multicolumn{2}{c}{(-2, 4) \quad (-3, 5)} \\
     \hline 
    \hline
    $(p_{2,-2}, r_{2,-2})$ & 
     \multicolumn{2}{c|}{(-5, 3) \quad (-4, 8)} & 
     \multicolumn{2}{c}{(-2, 2) \quad (-3, 2)} 
     \\
     \hline
    $(p_{2,-1}, r_{2,-1})$ & 
     \multicolumn{2}{c|}{(-4, 4) \quad (-6, 5)} & 
     \multicolumn{2}{c}{(-4, 3) \quad (-3, 2)} 
     \\
     \hline
    $(p_{2,0}, r_{2,0})$ & 
     \multicolumn{2}{c|}{(-4, 4) \quad (-4, 4)} & 
     \multicolumn{2}{c}{(-3, 3) \quad (-4, 3)} 
     \\
     \hline
    $(p_{2,1}, r_{2,1})$ & 
     \multicolumn{2}{c|}{(-4, 3) \quad (-3, 4)} & 
     \multicolumn{2}{c}{(-4, 2) \quad (-3, 3)}  
     \\
     \hline
    $(p_{2,2}, r_{2,2})$ & 
     \multicolumn{2}{c|}{(-3, 2) \quad (-5, 5)} & 
     \multicolumn{2}{c}{(-3, 2) \quad (-5, 2)}  
     \\
     \hline
     \hline
    \end{tabular}
  \end{center}
\end{table}

\begin{figure*}
    \centering
    \begin{minipage}{0.49\linewidth}
	\centering
	\includegraphics[width=0.9\linewidth]{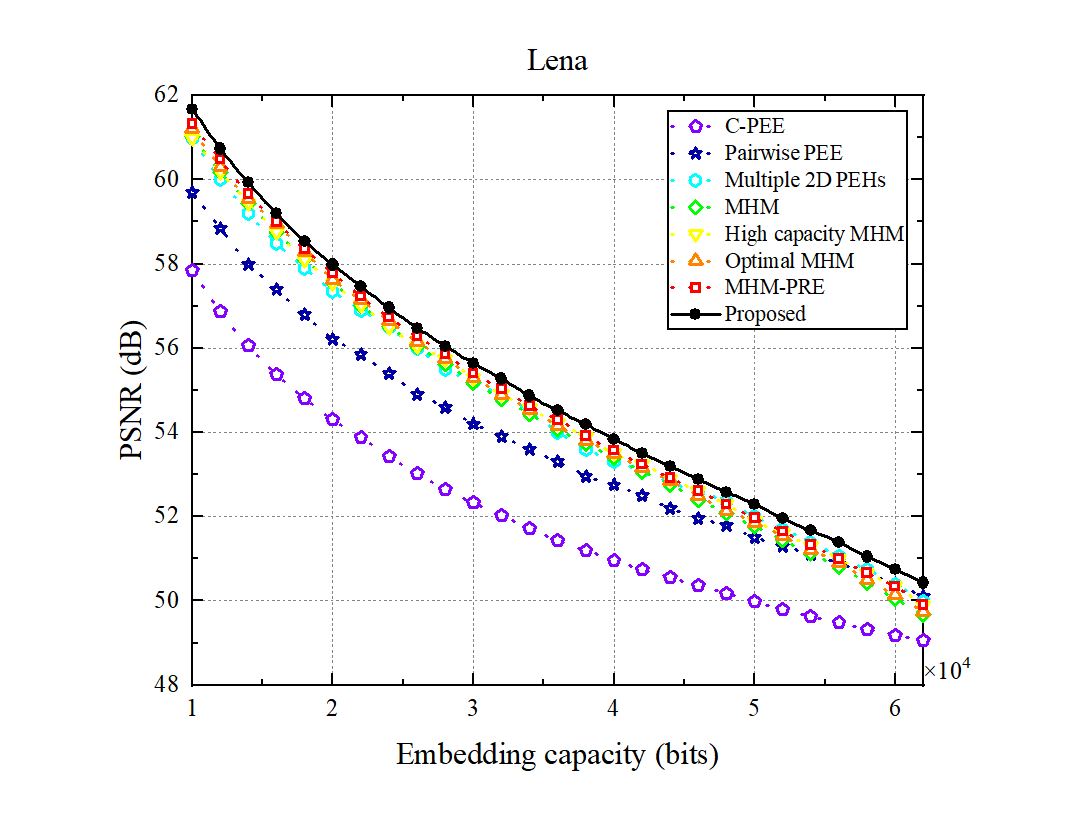}
    \end{minipage}
    \centering
    \begin{minipage}{0.49\linewidth}
	\centering
	\includegraphics[width=0.9\linewidth]{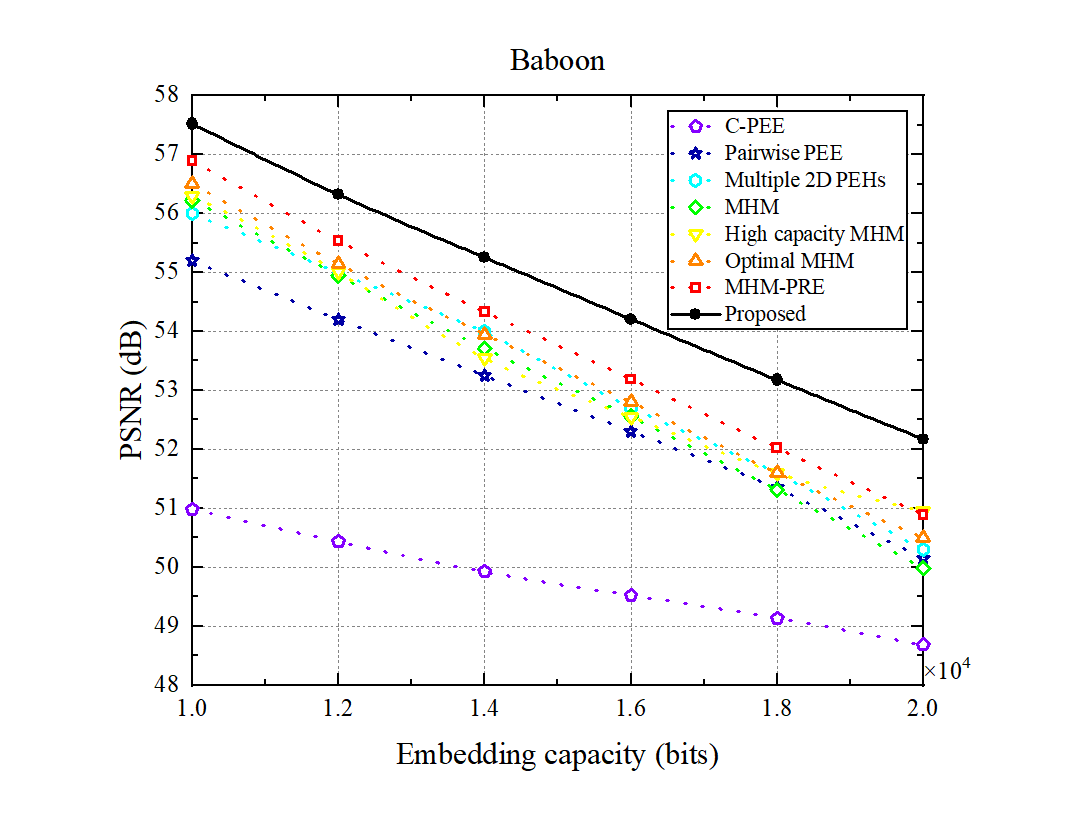}
    \end{minipage}

    \centering
    \begin{minipage}{0.49\linewidth}
	\centering
	\includegraphics[width=0.9\linewidth]{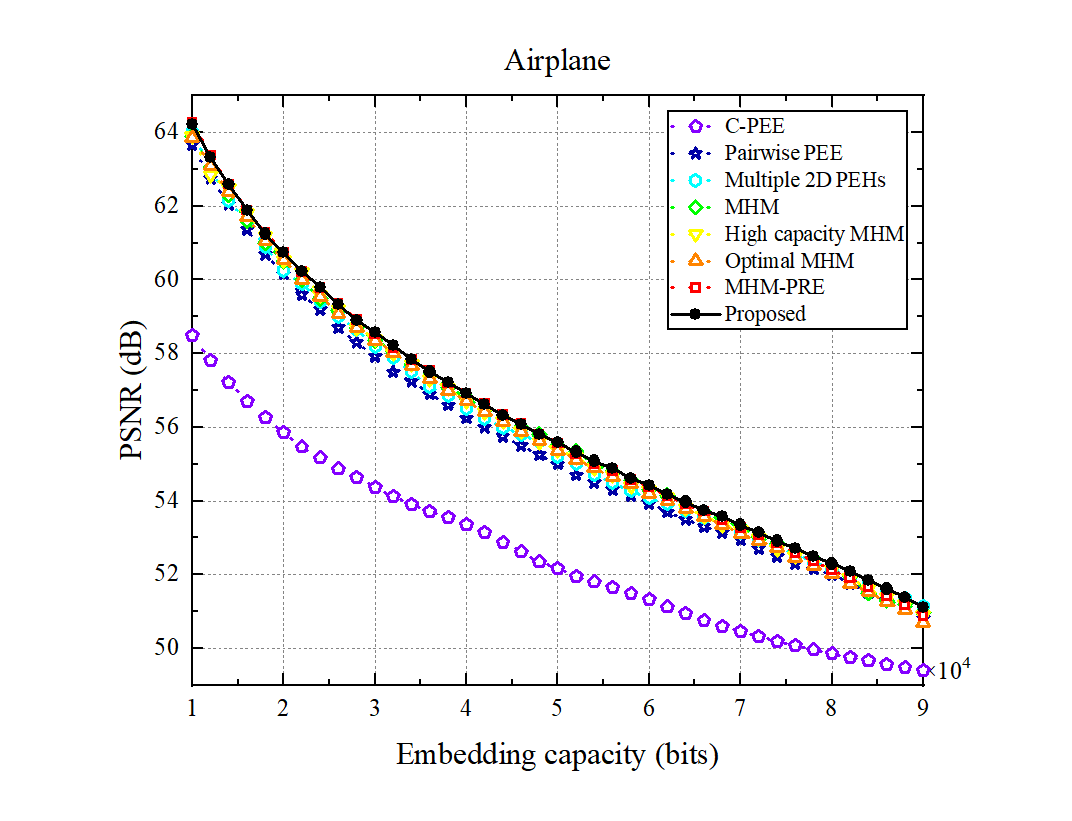}
    \end{minipage}
    \centering
    \begin{minipage}{0.49\linewidth}
	\centering
	\includegraphics[width=0.9\linewidth]{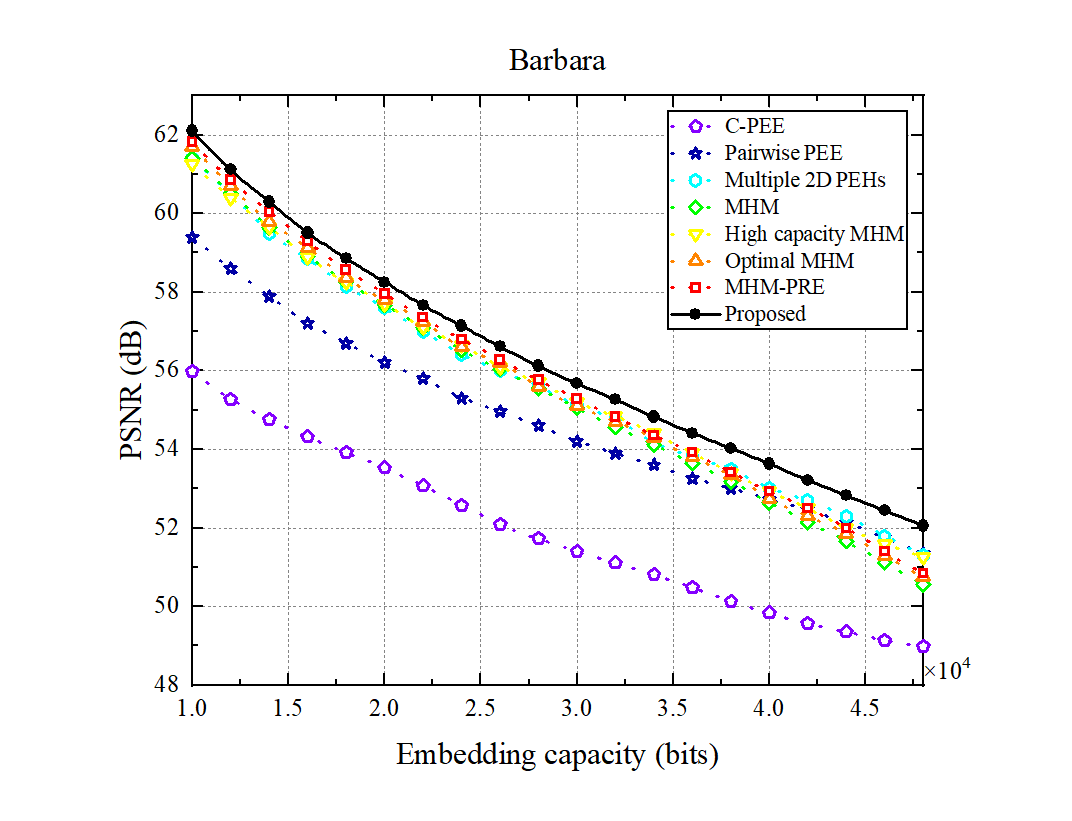}
    \end{minipage}

    \centering
    \begin{minipage}{0.49\linewidth}
	\centering
	\includegraphics[width=0.9\linewidth]{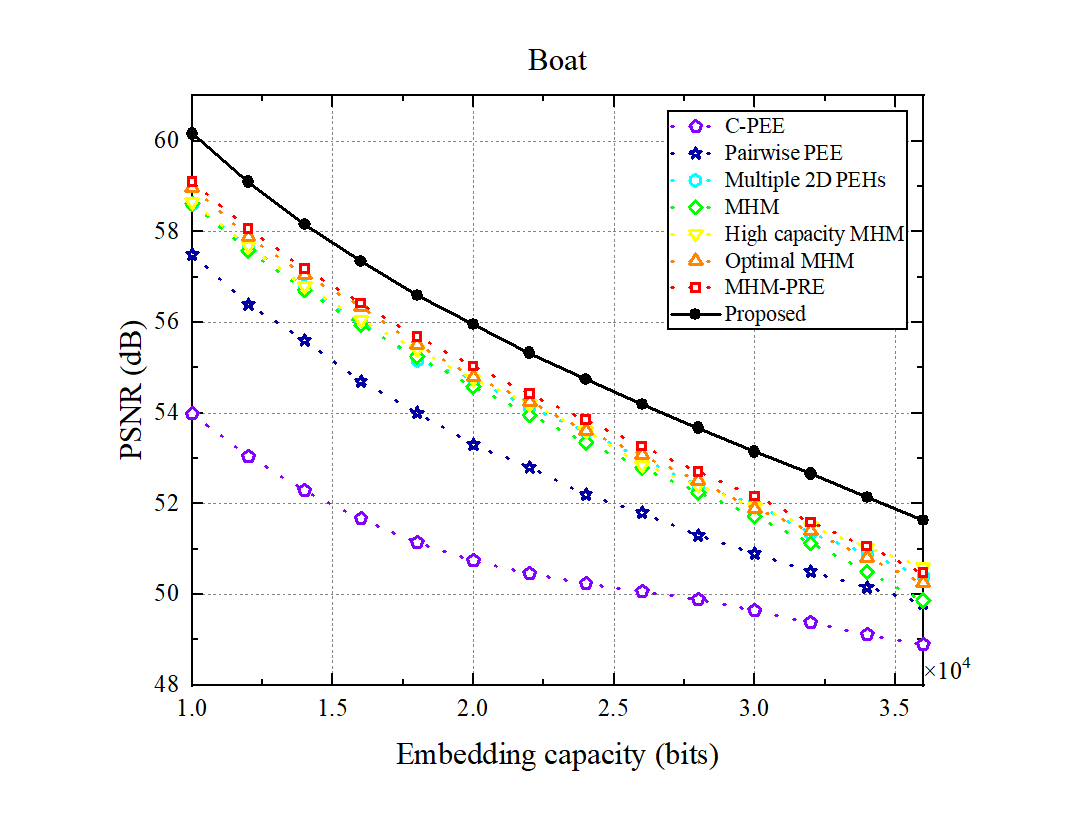}
    \end{minipage}
    \centering
    \begin{minipage}{0.49\linewidth}
	\centering
	\includegraphics[width=0.9\linewidth]{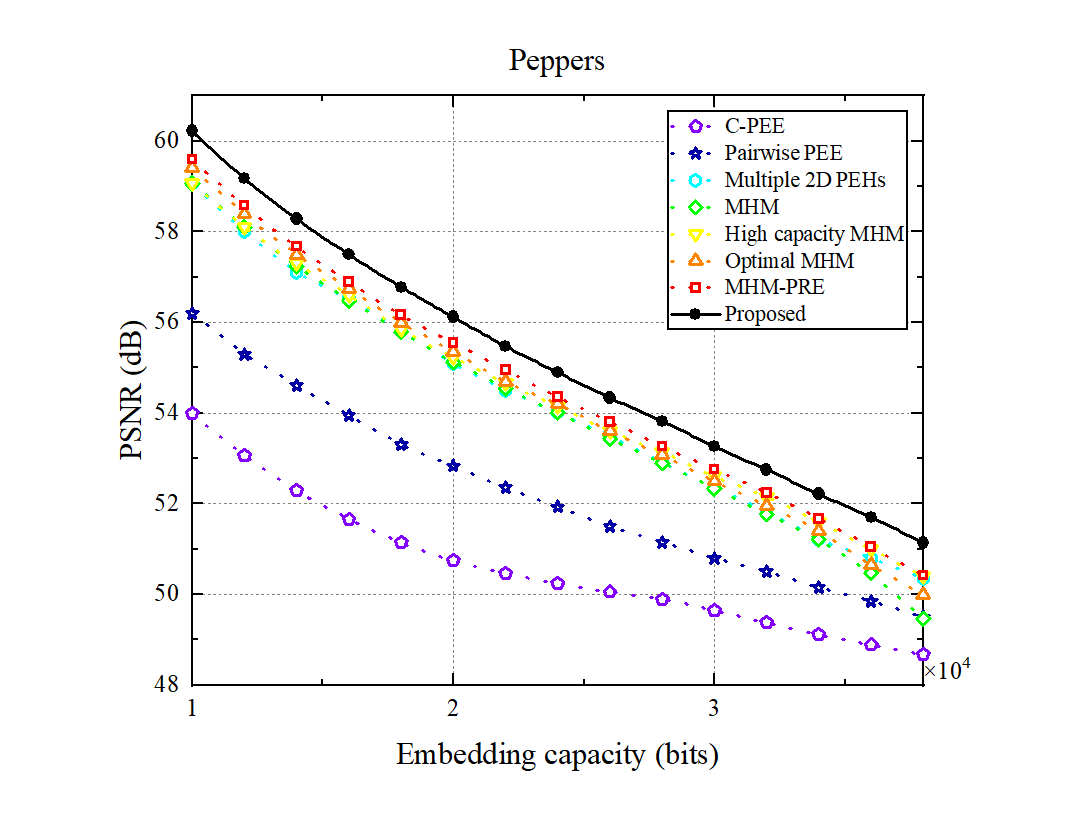}
    \end{minipage}
    \label{six_graphs}
    \caption{Performance comparison between the proposed method and C-PEE, Pairwise PEE, MHM, MHM-based methods}
\end{figure*}

For C-PEE, a very sharp 1D-PEH can be generated by rhombus predictor, double-layered embedding, has already achieved a good performance. However, from \ref{tabel_PSNR_1W} and \ref{tabel_PSNR_2W}, we can find that the average PSNR of the proposed mthod gains are 4.88 dB and 4.56 dB higher than C-PEE's in the embedding capacity of $10,000$ and $20,000$. For pairwise PEE proposed by Bo Ou et al., the capability of embedding is further improved by extending PEH to 2D and adaptively discarding the mapping direction with large distortion. From \ref{tabel_PSNR_1W} and \ref{tabel_PSNR_2W}, the proposed method outperforms pairwise PEE with an average PSNR increase of 2.33 dB and 2.10 dB in the embedding capacity of 10,000 bits and 20,000 bits respectively.

For original MHM proposed by Li Xiaolong et al., it initiatively proposes to generate series 1D-PEHs according to complexity thresholds, and select a pair of expansion bins in each PEH for embedding. From \ref{six_graphs}, we can see that MHM achieves good results on all kinds of images, and when the embedding capacity is $10,000$ and $20,000$ , the average PSNR reaches a very high level: 60.04dB and 55.91dB respectively. However, from \ref{tabel_PSNR_1W} and \ref{tabel_PSNR_2W}, the performance of the proposed method is still better than original MHM, improves it with an average PSNR gain of 0.95 dB and 0.96 dB. Since then, more and more researchers have conducted works based on MHM. Among them, the state-of-the-art works include High capacity MHM proposed by Bo Ou et al., in 2020, Optimal MHM proposed by Wenfa Qi et al., in 2020, and HMM-PRE proposed by Xiao Mengyao et al., in 2023. Unlike MHM, which selects only one pair of expansion bins, High capacity MHM selects multiple pairs of expansion bins for each PEH. Compared with other MHM-type algorithms, High capacity MHM not only achieves higher total embedding capacity, but also has a slower rate of performance degradation as the capacity increases. Optimal MHM optimizes MHM by continually continuously reducing the original problem to smaller subproblems. MHM-PRE replaces prediction-error with pixel-residual to establish multiple PRHs, and derives more expansion bins through a new mapping mechanism. From \ref{six_graphs}, although High capacity MHM, Optimal MHM and MHM-PRE perform well in these images, and both offer significant improvements over original MHM, the proposed method still outperforms them. It improves High capacity MHM with an average PSNR gain of 0.95 dB and 0.75 dB, improves Optimal MHM with an average PSNR gain of 0.79 dB and 0.76 dB, and improves MHM-PRE by 0.49 dB 与 0.54 dB PSNR gain in the embedding capacities of 10,000 bits and
20,000 bits, respectively. Moreover, on high-complexity image such as Baboon, the gain from double predictors will be more obvious, and the performance of the proposed method may perform even better.

To sum up, through comparison with C-PEE, Pairwise PEE, original MHM and several state-of-the-art methods based on MHM, the superiority and theoretical value of the proposed method is experimentally verified.

\section{Conclusion}
\label{Section V}

In this paper, we propose a improved 2D-PEH based on double prediction-error. First, different from previous 2D-PEH, the proposed 2D-DPEH is established by selecting two distinct predictors with low correlation to calculate double prediction-errors for each pixel. In addition, we adopt DP to optimize the selection of expansion bins, speeding up the running time and improving the quality of the embedded image. Finally, we combined the proposed method with C-PEE and original MHM, then designed comparative experiments with state-of-the-art Pee-based methods in recent years to verify the superiority of the proposed algorithm and extend PEE into a more general case.






\vspace{12pt}
\end{CJK}  
\end{document}